\documentclass[prd,reprint,nofootinbib,superscriptaddress]{revtex4-2}
\usepackage{graphicx,url,amssymb,amsmath,rotating,color,units,wasysym,epsfig,multirow,epstopdf, stackengine}
\usepackage{amsmath}
\usepackage{cancel}
\usepackage{physics}
\usepackage{appendix}
\usepackage{amssymb}
\usepackage[colorlinks,urlcolor=blue,citecolor=blue,linkcolor=blue]{hyperref}
\usepackage[normalem]{ulem}
\usepackage{tabularx}
\usepackage{tensor}
\newcolumntype{Y}{>{\centering\arraybackslash}X}

\usepackage{xcolor}

\renewcommand{\pop}{\rm pop}

\newcommand{\result}[1]{{\color{black} #1}}

\newcommand{\etagw}{\eta_{\rm GW}}
\newcommand{\pdet}{p_{\rm det}}
\newcommand{\eos}{\varepsilon}

\newcommand{\mpopgw}{M_{\rm pop, GW}}
\newcommand{\mpopem}{M_{\rm pop, EM}}
\DeclareMathOperator{\nuc}{nuc}
\DeclareMathOperator{\GW}{GW}
\DeclareMathOperator{\EM}{EM}
\DeclareMathOperator{\NICER}{NICER}
\DeclareMathOperator{\PSR}{PSR}

\newcommand{\posteriorMtov}{\result{2.28^{+0.41}_{-0.21}}}
\newcommand{\posteriorRtyp}{\result{12.2^{+0.8}_{-0.9}}}
\newcommand{\posteriorLambdatyp}{\result{438^{+224}_{-166}}}

\newcommand{\posteriormpopem}{\result{2.05^{+0.11}_{-0.06}}}
\newcommand{\posteriormpopgw}{\result{1.85^{+0.39}_{-0.16}}}
\newcommand{\popinformedjzerozeromass}{\result{1.37_{-0.11}^{+0.22}}}
\newcommand{\popinformedjzerosevenmass}{\result{2.01_{-0.09}^{+0.08}}}
\newcommand{\popinformedjzerofourmass}{\result{1.39_{-0.05}^{+0.08}}}
\newcommand{\posteriorcsmaxlowerbound}{\result{0.59}}
\newcommand{\posteriorrhocatmmax}{\result{{5.53}^{+1.07}_{-1.24}}}

\newcommand{\posteriormatfourrhonucupperbound}{\result{2.69}}

\newcommand{\posterioremmuone}{\result{{1.35}^{+0.02}_{-0.02}}}
\newcommand{\posterioremmutwo}{\result{{2.01}^{+0.43}_{-0.27}}}
\newcommand{\posterioremfracone}{\result{{0.65}^{+0.11}_{-0.13}}}
\newcommand{\posterioremsigmaone}{\result{{0.07}^{+0.02}_{-0.02}}}
\newcommand{\posterioremsigmatwo}{\result{{0.39}^{+0.37}_{-0.22}}}

\newcommand{\nojzerofourRtyp}{\result{{12.5}^{+1.0}_{-0.9}}}
\newcommand{\nojzerozeroRtyp}{\result{{11.6}^{+1.3}_{-0.9}}}
\newcommand{\nonicerRtyp}{\result{{11.9}^{+1.7}_{-1.6}}}

\newcommand{\nojzerofourcsmax}{\result{{0.8}^{+0.19}_{-0.31}}}
\newcommand{\legredcsmaxlowerbound}{0.51}
\newcommand{\legredmatfourrhonucupperbound}{\result{2.55}}

\newcommand{\mpopemmtovdiffninety}{\result{0.53\,M_\odot}}
\newcommand{\mpopgwmtovdiffninety}{\result{0.73\,M_\odot}}

\newcommand{\mpopgwmtovdiffninetylowspin}{\result{0.77\,M_\odot}}

\newcommand{\CIT}{\affiliation{Department of Physics, California Institute of Technology, Pasadena, California 91125, USA}}
\newcommand{\CITLab}{\affiliation{LIGO Laboratory, California Institute of Technology, Pasadena, California 91125, USA}}
\newcommand{\CITA}{\affiliation{Canadian Institute for Theoretical Astrophysics, University of Toronto, Toronto, Ontario M5S 3H8, Canada}}
\newcommand{\PI}{\affiliation{Perimeter Institute for Theoretical Physics, Waterloo, Ontario N2L 2Y5, Canada}}

\begin{document}
\title{The interplay of astrophysics and nuclear physics in determining the properties of neutron stars}

\author{Jacob Golomb}
\email{jgolomb@caltech.edu}
\CIT \CITLab

\author{Isaac Legred}
\email{ilegred@caltech.edu}
\CIT \CITLab

\author{Katerina Chatziioannou}
\email{kchatziioannou@caltech.edu}
\CIT \CITLab

\author{Philippe Landry}
\email{plandry@cita.utoronto.ca}
\CITA \PI

\begin{abstract}
Neutron star properties depend on both nuclear physics and astrophysical processes, and thus observations of neutron stars offer constraints on both large-scale astrophysics and the behavior of cold, dense matter.
In this study, we use astronomical data to jointly infer the universal equation of state of dense matter along with two distinct astrophysical populations: Galactic neutron stars observed electromagnetically and merging neutron stars in binaries observed with gravitational waves. 
We place constraints on neutron star properties and quantify the extent to which they are attributable to macrophysics or microphysics. 
We confirm previous results indicating that the Galactic and merging neutron stars have distinct mass distributions.
The inferred maximum mass of both Galactic neutron stars, $\mpopem=\posteriormpopem\,M_{\odot}$ (median and 90\% symmetric credible interval), and merging neutron star binaries, $\mpopgw=\posteriormpopgw\,M_{\odot}$, are consistent with the maximum mass of nonrotating neutron stars set by nuclear physics, $M_{\rm TOV} =\posteriorMtov\,M_\odot$.  
The radius of a $1.4\,M_{\odot}$ neutron star is $\posteriorRtyp\,$km, consistent with, though $\sim 20\%$ tighter than, previous results using an identical equation of state model.
Even though observed Galactic and merging neutron stars originate from populations with distinct properties, there is currently no evidence that astrophysical processes cannot produce neutron stars up to the maximum value imposed by nuclear physics. 
\end{abstract}

\maketitle

\section{Introduction}

The properties of neutron stars (NSs) depend on both the dense-matter physics that governs their interiors and the astrophysical context in which they form, evolve, and are observed~\cite{Lattimer:2015nhk,Ozel:2016oaf,Chatziioannou:2020pqz,Chatziioannou:2024tjq}.
This interplay is driven by an apparent  coincidence: the mass scale of maximally-compact matter in its ground state is comparable to the Chandrasekhar mass.
The NS characteristic compactness (defined as $M/R$ with $M$ the mass and $R$ its radius) is just below the black-hole (BH) limit of 1/2\footnote{In units where $G=c=1$, which we use unless otherwise stated.}.  
This implies  $2M/R  \sim c_s^2 \sim 1$~\cite{Saes:2024xmv}, where $c_s^2$ is the characteristic speed of sound squared in the body. 
In the standard model, cold matter can only achieve such sound-speeds at densities greater than an atomic nucleus, $\rho_{\rm nuc} \sim 2.8 \times 10^{14}\rm{g}/\rm{cm}^3$ at high neutron-to-proton ratio.  
This requirement fixes both the compactness \emph{and} density of such a near-maximally compact object, and therefore its mass and radius scales to $M \sim 1\,M_{\odot}$ and $R \sim 10\,\rm{km}$ respectively. 
The former is remarkably close to the Chandrasekhar mass, ${\sim} 1.4\,M_\odot$, the maximum mass that can be supported by electron degeneracy. 
As NSs form from cores that are too massive to be supported by electron degeneracy, this sets another characteristic mass scale for NSs~\cite{Couch:2017}. 

Substantial uncertainties in the details of NS formation and dense-matter physics mean it is not immediately clear which of the two drives the distribution of NS masses.
For example, general relativity and the dense-matter equation of state (EoS) set a maximum mass for nonrotating NSs, the Tolman-Oppenheimer-Volkoff (TOV) limit $M_{\rm TOV}$~\cite{Tolman:1939jz, Oppenheimer:1939ne}. Originally speculated to be near $0.7\,M_{\odot}$, $M_{\rm TOV}$ is now understood to be ${\sim} 2{-}3\, M_{\odot}$~\cite{Rhoades:1974fn, Kalogera:1996ci, Lattimer:2000nx, Pang:2021jta, Raaijmakers:2021uju, Legred:2021, Miller:2021qha}, but 
it is unknown whether astrophysical formation mechanisms can produce NSs up to this mass.   
Moreover, NSs form in a variety of ways, including in core-collapse supernovae and binary mergers, each of which likely results in different natal mass and spin distributions.  
Even after formation, NSs are modified via binary interactions: for instance, ``spider" pulsars~\cite{Roberts:2013} may achieve large masses and spins via accretion.

Galactic observations have constrained the masses of dozens of NSs in binaries via pulsar timing~\cite{Blanford:1976}. 
The mass distribution of Galactic NSs with a mass measurement includes a primary peak at ${\sim}1.35\, M_{\odot}$ preferred at $3{:}1$ over a secondary peak at ${\sim} 1.8\, M_{\odot}$~\cite{Antoniadis:2016hxz,Alsing:2017bbc,Farr_2019}.  
The observed cutoff in the distribution above ${\sim}2\,M_{\odot}$~\cite{Alsing:2017bbc,Farr_2019} may correspond to the TOV mass, or to a different maximum mass imposed by astrophysical processes; the most general interpretation of the cutoff identifies it as an astrophysical maximum mass that may differ from $M_{\rm TOV}$.
The Galactic NS population is broadly consistent with the masses of NS-like compact objects in wide-period binaries revealed by Gaia astrometry~\cite{El-Badry:2024a, El-Badry:2024b}.
However, this inferred mass distribution does not account for selection effects in the various surveys, and lumps together NSs in different astrophysical systems, e.g., NS--NS binaries (BNS) and NS--WD binaries, that may have different inherent distributions.
Indeed, the known Galactic BNSs are all contained within the ${\sim}1.35\, M_{\odot}$ component of the bimodal distribution~\cite{Farrow:2019xnc}.

A subset of Galactic millisecond pulsars~\cite{Manchester:2017azn} show persistent pulsed X-ray emission originating from surface hotspots. 
Detailed modeling of the hotspot emission has placed constraints on the mass and radius of three pulsars using NICER and XMM-Newton~\cite{Miller:2019cac, Riley:2019yda, Miller:2021qha, Riley:2021pdl, Choudhury:2024xbk}, two of which are in binaries and thus have radio-based mass constraints. 
Since two of the NICER targets are known radio pulsars, they are commonly treated as part of the Galactic NS population.  
For example, the properties of PSR J0740+6620, one of the most massive known pulsars~\cite{Cromartie:2019kug,Fonseca:2021wxt}, have been inferred simultaneously with the Galactic population~\cite{FarrChatziioannou2020}.  
Requiring PSR J0740+6620 to hail from the bimodal Galactic NS mass distribution revises its mass downward to $2.03^{+0.14}_{-0.11}\,M_{\odot}$~\cite{FarrChatziioannou2020}.

A different population consists of NSs in merging compact binaries with NSs or black holes (BHs) observed with gravitational waves (GWs)~\cite{KAGRA:2021duu}.
Among BNSs, GW170817~\cite{gw170817} is consistent with the Galactic BNS population with a total mass of ${\sim}2.7\,M_{\odot}$.
GW190425, at a total mass of ${\sim} 3.4\, M_{\odot}$~\cite{gw190425}, is however an outlier.
Attempts to explain this discrepancy include selection effects~\cite{Romero-Shaw:2020aaj, Safarzadeh:2020efa} and non-BNS interpretations~\cite{Han:2020qmn, Foley:2020kus}.
Regardless, this discrepancy suggests that the Galactic and merging BNS distributions should be treated separately.
The distribution of all NSs observed in merging binaries to date, including both BNSs and likely NSBH systems~\cite{LIGOScientific:2021qlt, LIGOScientific:2024elc}, is relatively flat with no prominent peak at ${\sim}1.35\,M_{\odot}$~\cite{Landry:2021hvl,KAGRA:2021duu}. The population of NSs in BNSs and NSBHs might, however, be different owing to different formation and evolutionary histories~\cite{KAGRA:2021duu, Biscoveanu:2022iue}. 
NS spins are ignored from these constraints due to large measurement uncertainties~\cite{Biscoveanu:2021eht}; it is therefore unknown how merging NS spins relate to the well-measured spins of Galactic NSs.
GW-based NS observations (primarily GW170817) also drive constraints on the EoS through mass and tidal deformability constraints~\cite{gw170817, Abbott:2018exr,Abbott:2018wiz}.  

The picture is much simpler when it comes to the nuclear physics and the EoS of NSs. 
Even when originating from different formation mechanisms, cold NSs are expected to be described by the same universal EoS. 
This expectation has been widely utilized to combine mass, radius, and tidal deformability measurements from various observations to place constraints on the EoS, e.g.~\cite{Abbott:2018exr,  Miller:2019nzo,Raaijmakers:2019dks, Dietrich:2020efo, Landry:2020vaw, Greif:2020pju, Miller:2021qha,Raaijmakers:2021uju,  Pang:2021jta,  Legred:2021,  Fan:2023spm, Biswas:2024hja}.
Even then, assumptions about NS masses have to be made.

Such assumptions typically include a uniform mass distribution, and whether astrophysical mechanisms create NSs up to the TOV mass or up to a different predetermined value~\cite{Landry:2020vaw,Legred:2021}.

In this paper, we study the properties of NSs in binaries with a focus on separating the impact of nuclear physics and astrophysics.
We use radio, X-ray, and GW data to jointly infer the dense-matter EoS and the NS mass distribution, each with their own maximum mass.
We go beyond considering a single mass distribution for all NSs that terminates at the TOV mass~\cite{Biswas:2024hja,Fan:2023spm} and separately infer the populations of Galactic NSs and merging BNSs.
Moreover, rather than the TOV mass, we allow the possibility of the astrophysical mass distribution terminating at a different ``astrophysical maximum mass" that is lower than the TOV mass.
Our model and inference set up allow us to begin to answer whether the maximum mass of NSs in various subpopulations is limited by the EoS or by astrophysical processes.
Beyond access to such questions, simultaneous inference mitigates biases that can arise with as few as $\mathcal{O}(10)$ GW detections when inferring either the EoS  or the mass distribution alone while making improper assumptions about the other~\cite{Wysocki:2020myz, Golomb:2021tll}. 
We also account for GW selection effects, which cause the detected population to be biased towards higher masses; as the selection effects in the electromagnetic surveys are unknown, we do not consider them.

The subpopulations, datasets, and models are described in Sec.~\ref{sec:population-models}.
The EoS is modeled with a mixture of Gaussian processes (GPs)~\cite{Landry:2018prl, Essick:2019ldf}, which allows for a wide range of EoS morphologies including phase transitions~\cite{Essick:2023fso} and imposes minimal intra-density correlations that hamper the flexibility of parametric models~\cite{Legred:2022pyp}.
We consider two subpopulations:
\begin{enumerate}
    \item The Galactic NS population is modeled with a bimodal distribution with a maximum mass cutoff~\cite{Alsing:2017bbc, Antoniadis:2016hxz}. The relevant datasets include radio, optical, and X-ray observations of pulsars in binaries~\cite{Alsing:2017bbc} and X-ray pulse-profile observations of pulsars J0030+0451~\cite{Miller:2019cac, Riley:2019yda}, J0740+6620~\cite{Miller:2021qha, Riley:2021pdl}, and J0437-4715~\cite{Choudhury:2024xbk}.
    \item The merging BNS population observed with GWs is modeled with a power-law with a maximum mass cutoff. The dataset consists of GW170817~\cite{gw170817} and GW190425~\cite{gw190425}, both assumed to be BNSs.
\end{enumerate}
Joint inference on the EoS and mass subpopulations is performed with a reweighting scheme that is described in Sec.~\ref{sec:joint-inference} and Appendix~\ref{app:likelihood}.

Our results are presented in Sec.~\ref{sec:results}. 
We find no evidence that the maximum mass of the two subpopulations is different than the TOV mass.
The difference between the maximum Galactic NS (merging BNS) mass and the TOV mass is less than $\mpopemmtovdiffninety$ ($\mpopgwmtovdiffninety$) at 90\% credibility, with zero difference consistent with the posteriors. 
Even though the maximum masses are consistent, we confirm previous results that the mass distributions of Galactic NSs and merging BNSs are different.
The latter possesses no prominent peak at $1.35\,M_{\odot}$, indicating that the two distributions should be modeled separately in an inference framework and have the freedom to differ from one another.

For the NS EoS, we infer a sound-speed profile that exceeds the conformal bound of $1/\sqrt{3}$ for weakly interacting nucleonic matter~\cite{BedaqueSteiner2015}, in line with a previous study using the same EoS model~\cite{Legred:2021}: the 90\% lower bound on the maximum speed of sound squared anywhere inside the NS is $\posteriorcsmaxlowerbound$. 
We constrain the radius of a canonical NS, a proxy for the stiffness of the EoS, to $R_{1.4}=\posteriorRtyp\,\rm{km}$, and the TOV mass to $M_{\rm TOV}=\posteriorMtov \,M_{\odot}$. 
Uncertainties are lower than \citet{Legred:2021} due to the recent NICER observation of PSR J0437-4715 and the impact of the ensemble of Galactic NS mass measurements via the updated treatment of the maximum mass.

We conclude in Sec.~\ref{sec:conclusions}.

\section{Modeling the Equation of State and the mass distribution}
\label{sec:population-models}

In this section we describe the data, as well as the EoS and astrophysical populations we model them with.

\subsection{Data} \label{sec:data}

The observations that inform the joint inference of the NS EoS and astrophysical population come from three sources: radio/optical pulsar mass measurements (PSR), X-ray pulse profile modeling for pulsar masses and radii (NICER), and GW constraints on BNS masses and tidal deformabilities (GW).

The PSR dataset includes the 74 Galactic pulsars with a mass measurement from Ref.~\cite{Alsing:2017bbc}, minus PSR J0437--4715, which is counted as part of the NICER dataset.\footnote{While J0437--4715 is in the NICER dataset, we use its radio mass measurement to inform the mass distribution.} The PSR observations are heterogeneous, including NSs in various types of binaries and several different mass determination methods.

The NICER dataset consists of the observations of PSR J0030+0451~\cite{Miller:2019nzo, Riley:2019yda}, PSR J0740+6620~\cite{Miller:2021qha, Riley:2021pdl}, and PSR J0437--4715~\cite{Choudhury:2024xbk}. 
The constraints on the masses and radii of these pulsars are sensitive to the details of the X-ray pulse profile modeling, such as the assumed hotspot geometry and the stochastic sampling of the multidimensional parameter posterior; thus different interpretations of the NICER data exist. Here we use results from the three-hotspot 
model of Ref.~\cite{Miller:2019nzo} for J0030+0451, the combined NICER-XMM Newton analysis with the two-hotspot model from Ref.~\cite{Miller:2021qha} for J0740+6620, and the \texttt{CST+PDT} model from Ref.~\cite{Choudhury:2024xbk} for J0437--4715. 
As the NICER analyses for J0740+6620 and J0437--4715 incorporate pre-existing radio-based mass estimates, we exclude them from the PSR dataset to avoid double-counting. In Appendix~\ref{sec:no-nicer} we quantify the sensitivity of our inference to alternative data selection choices for the NICER observations.

For the GW dataset, we consider compact binary coalescences from the third Gravitational Wave Transient Catalog~\cite{KAGRA:2021vkt} of the LIGO-Virgo-KAGRA network~\citep{TheLIGOScientific:2014jea, TheVirgo:2014hva, KAGRA:2020tym} with source-frame chirp mass $\mathcal{M} \lesssim 2.176 \,M_{\odot}$, corresponding to equal-mass component masses below $2.5 \,M_{\odot}$.
This leaves us with GW170817~\cite{gw170817} and GW190425~\cite{gw190425} as the only events consistent with BNS mergers.
We do not consider the recent observation of $\rm GW230529\_181500$~\cite{LIGOScientific:2024elc}, which is potentially a BNS merger according to this criterion, as sensitivity estimates for the fourth observing run do not exist. 
For GW170817, we generate new posterior samples with the waveform approximant \texttt{IMRPhenomPv2\_NRTidal}, which includes spin-precession and tidal effects~\citep{Dietrich:2018uni}, using the parameter estimation package \texttt{bilby}~\citep{Ashton2019, Romero-Shaw:2020owr} and the nested sampler \texttt{dynesty}~\citep{Speagle20}. 
We fix the source location to the host galaxy NGC4993 and adopt spin priors that are isotropic in orientation and uniform in dimensionless magnitude up to $0.05$, motivated by the spin distribution of pulsars in binary systems expected to merge within a Hubble time~\citep{Zhu:2017znf}.
For GW190425, we use the publicly released parameter estimation samples~\cite{190425samples} for the \texttt{IMRPhenomPv2\_NRTidal} waveform. 
Since GW190425's total mass is inconsistent with those of Galactic BNSs, we allow for dimensionless spin magnitudes up to 0.4, roughly corresponding to a $1\,$ms spin period~\cite{Hessels2006}. Appendix~\ref{app:lowspin} investigates the impact of a spin-magnitude upper limit of 0.05 for both GW170817 and GW190425. 

\subsection{EoS model}

The dense-matter EoS, i.e., the pressure-density relation, is described with a model-agnostic Gaussian process~\cite{Landry:2018prl, Essick:2019ldf}, which builds a prior EoS process via a mixture of GP hyperparameters probing a large range of correlation scales and strengths.  
This procedure produces an EoS distribution that is relatively insensitive to the nuclear models it is conditioned on~\cite{Essick:2019ldf} and imposes minimal model-dependent correlations between the low- and high-density EoS~\cite{Legred:2022pyp}.  
The GP flexibility is particularly important for our goal of disentangling the maximum TOV mass $M_{\rm TOV}$ and the maximum astrophysical mass.
Less flexible parametric EoS models implicitly correlate the radius or tidal deformability and $M_{\rm TOV}$~\cite{Legred:2022pyp} which in turn translate to model-dependent correlations between $M_{\rm TOV}$ and the astrophysical parameters.  
All NSs are assumed to be described by the same EoS.
For efficiency, we restrict the prior to EoSs with $M_{\rm TOV} > 1.8\,M_{\odot}$. 

\subsection{Astrophysical population models}

For the astrophysical mass distribution we use parametric distributions with hyperparameters $\eta$. 
We consider two classes of observations modeled with separate distributions: Galactic NSs observed via electromagnetic (EM) radiation as part of the PSR and NICER datasets, and NSs in merging BNSs observed via GWs constituting the GW dataset. 

We restrict to the NS masses while ignoring spins and assume that all objects are NSs.

\subsubsection{Galactic neutron stars with radio and X-rays}

Motivated by Refs.~\cite{Antoniadis:2016hxz,Alsing:2017bbc,FarrChatziioannou2020}, we model the Galactic NS masses $m$ as a mixture of two Gaussians:
\begin{equation}
\pi(m|\eta_{\EM}) = 
\label{eq:pulsar-population}
      f \mathcal{N}(\mu_1, \sigma_1) + (1 - f) \mathcal{N}(\mu_2, \sigma_2)\,, 
\end{equation}
for $m \in [M_{\min}, \mpopem]$, and where $\mathcal{N}(\mu, \sigma)$ is a truncated normal distribution with mean $\mu$ and standard deviation $\sigma$, and $f$ is the mixture weight. Following Ref.~\cite{Alsing:2017bbc}, we fix $M_{\min} = 1 \,M_\odot$ and infer the hyperparameters $\eta_{\EM}=\{\mu_1,\mu_2,\sigma_1,\sigma_2,f,\mpopem\}$ with flat priors: $\mpopem \in (1.8, 3.0) \,M_\odot$, $\mu_1 \in (1, 2)\, M_\odot$, $\mu_2 \in (\mu_1, 2.5) \,M_\odot$, $f \in (0, 1)$, and $\sigma_{1,2} \in (0.05, 1) \,M_\odot$.
Since all analyzed objects are NSs, we impose $M_{\rm pop,EM} < M_{\rm TOV}$.\footnote{We ignore the impact of pulsar spin on the maximum mass. Using approximate relations to fourth order in spin magnitude~\cite{Most:2020bba, Breu:2016ufb}, we estimate that the maximum allowed mass will differ from $M_{\rm TOV}$ by $\lesssim 1\%$ compared to statistical uncertainties ${\sim}20-30\%$ for the range of pulsar periods in our dataset, $ P \gtrsim 2\, \rm{ms}$.}  
This prior restriction leads to a marginal priors on $\mpopem$ and the EoSs that are not uniform, although the full multidimensional prior is flat within its domain of support.

Although the PSR and NICER datasets include NSs in different astrophysical settings, i.e.~in binaries with various companions, or in isolation in the case of J0030+0451, and could in principle hail from different subpopulations, we model these NSs as a single population for consistency with previous results and due to the lack of selection effect estimates. (We are not aware of any established methods to account for selection effects in radio surveys or for NICER's target selection procedure~\cite{Bogdanov:2019ixe}.) 
Given the lack of selection effect estimates for the PSR and NICER datasets, we simply assume the observed mass distribution to be equivalent to the astrophysical distribution.\footnote{This procedure can result in a bias even for the detected population~\cite{Essick:2023upv}. Such a bias however is expected to be small.
For example, Fig.~4 of~\cite{Essick:2023upv} shows the bias for ${\sim} 800$ simulated GW observations. } We quantify the impact of this assumption in Appendix~\ref{app:uniform_em_population}, where we present results with a fixed uniform mass distribution in place of Eq.~\eqref{eq:pulsar-population}.

\subsubsection{Merging neutron stars with gravitational waves}
\label{sec:gw-observations}

We model BNS masses with a truncated power-law for both binary components $m_1$ and $m_2$:
\begin{equation}
    \label{eq:gw-population}
    \pi(m_1,m_2|\eta_{\GW}) \propto 
      m_1^\alpha m_2^\alpha \,,
\end{equation}
for $m \in [m_{\min}, \mpopgw]$ and random pairing between $m_1$ and $m_2$ in the two-dimensional space.
We again fix $m_{\min}=1\,M_{\odot}$ and infer the hyperparameters $\eta_{\GW}=\{\alpha,\mpopgw\}$ with flat priors $\alpha \in (-5, 5)$, $\mpopgw \in (1.6, 2.5)\,M_\odot$. 
Since we assume that both GW170817 and GW190425 are BNSs, we again impose $M_{\rm pop,GW} < M_{\rm TOV}$.

GW selection effects are well understood, and we incorporate them in our inference. 
Because the GW data selection procedure involves identifying events as BNSs based on a component mass cut at $2.5\,M_{\odot}$, our analysis only places constraints on the mass distribution below $2.5\,M_{\odot}$. The GW selection modeling is described in Sec.~\ref{gw-likelihood}. 

\section{Joint inference via reweighting}
\label{sec:joint-inference}

The joint mass-EoS model is a combination of EoS draws from the GP prior process and the parametric mass models of Eqs.~\eqref{eq:pulsar-population} and~\eqref{eq:gw-population}.
While the joint posterior could be sampled with standard stochastic sampling methods with pre-computed GP draws, we instead use a multi-stage reweighting scheme and the GP draws from Ref.~\cite{Legred:2021}.

The reweighting scheme includes the following steps, with technical details relegated to the Appendices:
\begin{enumerate}
    \item Use standard stochastic sampling to infer the mass population and the EoS using Eqs.~\eqref{eq:pulsar-population} and~\eqref{eq:gw-population} for the mass distribution and a simplified, low-dimensional EoS model. Details about the EoS model are given in Appendix~\ref{app:EoS-simple}.
    The EoS model is included here to mitigate potential biases of a mass-only inference~\cite{Golomb:2021tll}.
    \item Treat the inferred mass distribution as a proposal distribution. For each sample from the distribution of $\eta=\{\eta_{\EM},\eta_{\GW}\}$, calculate the likelihood for each pre-computed GP draw. The likelihood form depends on the dataset considered~\cite{Landry:2021hvl} and is described in Secs.~\ref{gw-likelihood} and~\ref{em-likelihood} for the GW and EM data respectively.
    \item With these likelihoods, calculate weights from the proposal mass distribution to the target joint mass-GP EoS distribution as described in Appendix~\ref{app:likelihood}.
    \item Combine the new posterior distributions for each dataset. This procedure allows us to obtain weighted samples from the joint posterior of the mass distribution and the GP EoS. We validate the reweighting scheme in Appendix~\ref{app:validation} with simulated GW observations.
\end{enumerate}

Each of the datasets considered (GW, NICER, and PSR) results in unique constraints and thus requires a unique formulation of the likelihood~\cite{Landry:2021hvl,Chatziioannou:2020pqz}. 
Below we discuss each dataset likelihood noting that the full likelihood is the product over the individual datasets.

\subsection{GW likelihood}
\label{gw-likelihood}

Given $N_{\rm GW}$ independent events, the likelihood for the EoS $\eos$ and population hyperparameters $\eta_{\GW}$ is\footnote{This expression assumes a $1/R$ prior on the event rate $R$ and marginalizes over it~\citep{Mandel:2018mve, Fishbach:2018edt}.} \cite{Vitale:2020aaz, Thrane_2019, Mandel:2018mve}
\begin{equation}\label{eq:GWlikelihood}
\begin{split}
    {\cal{L}}_{\GW}&(d|\eos, \eta_{\GW}) \propto \ \pdet (\etagw)^{-N_{\GW}} \times \\ &\prod^{N_{\GW}}_i \int {\cal{L}}(d_i|m_1, m_2, \eos) \pi(m_1, m_2| \eta_{\GW})dm_1 dm_2\,,
\end{split}
\end{equation}
where $\pi(m_1, m_2|\eta_{\GW})$ is the model of Eq.~\eqref{eq:gw-population} and 
\begin{equation}
    {\cal{L}}(d_i | m_1, m_2, \eos) = \mathcal L(d_i | m_1, m_2, \Lambda(m_1, \eos)\Lambda(m_2, \eos))\,,
\end{equation} 
is the $i$th individual-event GW likelihood (e.g., \cite{Finn:1992wt, Veitch:2014wba}) marginalized over all binary parameters other than the component masses $m_1,m_2$ and tidal deformabilities $\Lambda_1,\Lambda_2$. 
Consistency with the EoS is ensured by calculating the likelihood for $\Lambda_1=\Lambda(m_1, \eos), \Lambda_2=\Lambda(m_2, \eos)$, i.e., the EoS prediction for the tidal deformability given the mass.
We estimate the individual-event likelihood from the posterior samples for the source-frame masses and tidal deformabilities using a Gaussian mixture model~\cite{Golomb:2021tll}, and the integral in Eq.~\eqref{eq:GWlikelihood} is computed as a Monte Carlo sum.  

The term $p_{\rm det}(\eta_{\rm GW})$ 
encodes the selection effect which characterizes how parts of the parameter space are over-represented in a catalog of GW events, as determined by the sensitivity of the detectors.
Defining $p_{\rm det}(d)$ as the probability that search algorithms detect a significant signal in data $d$ results in
\begin{equation}\label{eq:pdet}
\begin{split}
        p_{\rm det}(\eta_{\GW}) \equiv \int \mathcal Dd & \int d\theta\,  p(d|\theta) \pi(\theta|\eta_{\GW}) p_{\rm det}(d)\\
        = & \int d \theta\, \pi(\theta|\eta_{\GW}) p_{\rm det}(\theta)\,,
\end{split}
\end{equation}
where we identify $p_{\rm det}(\theta) \equiv \int \mathcal D d\, p(d|\theta) p_{\rm det}(d)$ as the probability of detecting an event with parameters $\theta$, marginalized over possible realizations of data $d$.       
For example, neglecting the specifics of the noise-generating process, the sensitivity to an event increases with its chirp mass ${\sim}\mathcal M_c^{5/6}$ and decreases inversely with its distance. 
We then further marginalize over possible realizations from the population $\theta \sim \pi(\theta|\eta_{\GW})$.
The presence of $p_{\rm det}(\eta_{\rm GW})$ in Eq.~\eqref{eq:GWlikelihood} ensures that the final result reflects the true astrophysical population rather than the observed population.
In practice, $\pdet(\eta_{\GW})$ might also depend on the EoS, but Ref.~\cite{Cullen:2017oaz} showed that the effect is negligible except for very stiff EoSs and low-mass NSs: there is a $\lesssim 2 \%$ change in the match between a template that sets $\Lambda=0$ and the true waveform. 

We compute $\pdet(\eta_{\GW})$ by reweighting recovered simulated signals in data from the first three observing runs, using standard techniques~\citep{KAGRA:2021duu, Farr_2019, Tiwari:2017ndi}.

\subsection{NICER likelihood}
\label{em-likelihood}

Given $N_{\rm NICER}$ observations, the likelihood for the EoS $\eos$ and population hyperparameters $\eta_{\EM}$ is obtained by marginalizing over the pulsar mass 
\begin{equation}\label{eq:NICERlikelihood}
    p_{\rm NICER}(d|\eos, \eta_{\EM}) = \prod^{N_{\NICER}}_i \int \mathcal{L}(d_i| m, \eos) \pi(m| \eta_{\EM}) dm\,,
\end{equation}
where $i$ indexes the NICER observations,  $\pi(m| \eta_{\EM})$ is the mass distribution of Eq.~\eqref{eq:pulsar-population}, and 
\begin{equation} \label{eq:NICERlikelihood2}
    \mathcal{L}(d_i| m, \eos)=\mathcal{L}(d_i| m, C(m,\eos))\,,
\end{equation}
is the individual-pulsar likelihood marginalized over all NICER parameters other than the mass $m$ and compactness $C$, which is again evaluated on the EoS prediction. The likelihoods are described in the publications associated with each observation \citep{Choudhury:2024xbk, Miller:2019cac, Miller:2021qha}.
We use a Gaussian mixture model~\cite{Golomb:2021tll} to evaluate Eq.~\eqref{eq:NICERlikelihood2}, and a Monte Carlo sum for the integral in Eq.~\eqref{eq:NICERlikelihood}. 

The NICER analysis of PSR J0437-4715 in Ref.~\cite{Choudhury:2024xbk} uses a prior that is flat in radius, rather than flat in compactness (or inverse compactness) like the analyses of PSR J0030-0451-~\cite{Miller:2019cac} and PSR J0770+6620~\cite{Miller:2021qha}. We correct for this with the appropriate Jacobian term to obtain a likelihood function in mass and compactness. 
Unlike Eq.~\eqref{eq:GWlikelihood} for the GW observations, the NICER likelihood ignores selection effects per the discussion in Sec.~\ref{sec:data}.

\subsection{PSR likelihood}
\label{psr-likelihood}

Finally, the likelihood for $N_{\PSR}$ pulsar mass measurements is
\begin{equation}
\label{eq:psrlikelihood}
    p_{\rm PSR}(d|\eos, \eta_{\EM}) = \prod^{N_{\PSR}}_i \int \mathcal{L}(d_i| m) \pi(m| \eta_{\EM}) dm\,,
\end{equation}
where $i$ indexes the pulsars and  $\pi(m| \eta_{\EM})$ is the mass distribution of Eq.~\eqref{eq:pulsar-population}.
The form of $\mathcal{L}(d_i|m)$ for each observation is prescribed analytically in Refs.~\cite{Alsing:2017bbc,FarrChatziioannou2020}, depending on whether the measurement constrains the pulsar mass, the binary mass function and the total mass, or the binary mass function and the mass ratio.
Like the NICER likelihoods, the PSR likelihoods do not account for selection effects, and we evaluate the integral in Eq.~\eqref{eq:psrlikelihood} via Monte Carlo.

\section{Implications of joint mass-EoS inference}
\label{sec:results}

In this section, we present results from the joint inference over the EoS and the mass distribution of two NS populations.
We begin with mass-specific and EoS-specific results in Secs.~\ref{sec:astro} and~\ref{sec:eos} respectively, before contrasting their impact on NS properties in Sec.~\ref{sec:joint}.

\subsection{Constraints on astrophysical populations}
\label{sec:astro}

\begin{figure}
    \centering  
    \includegraphics[width=.46\textwidth]{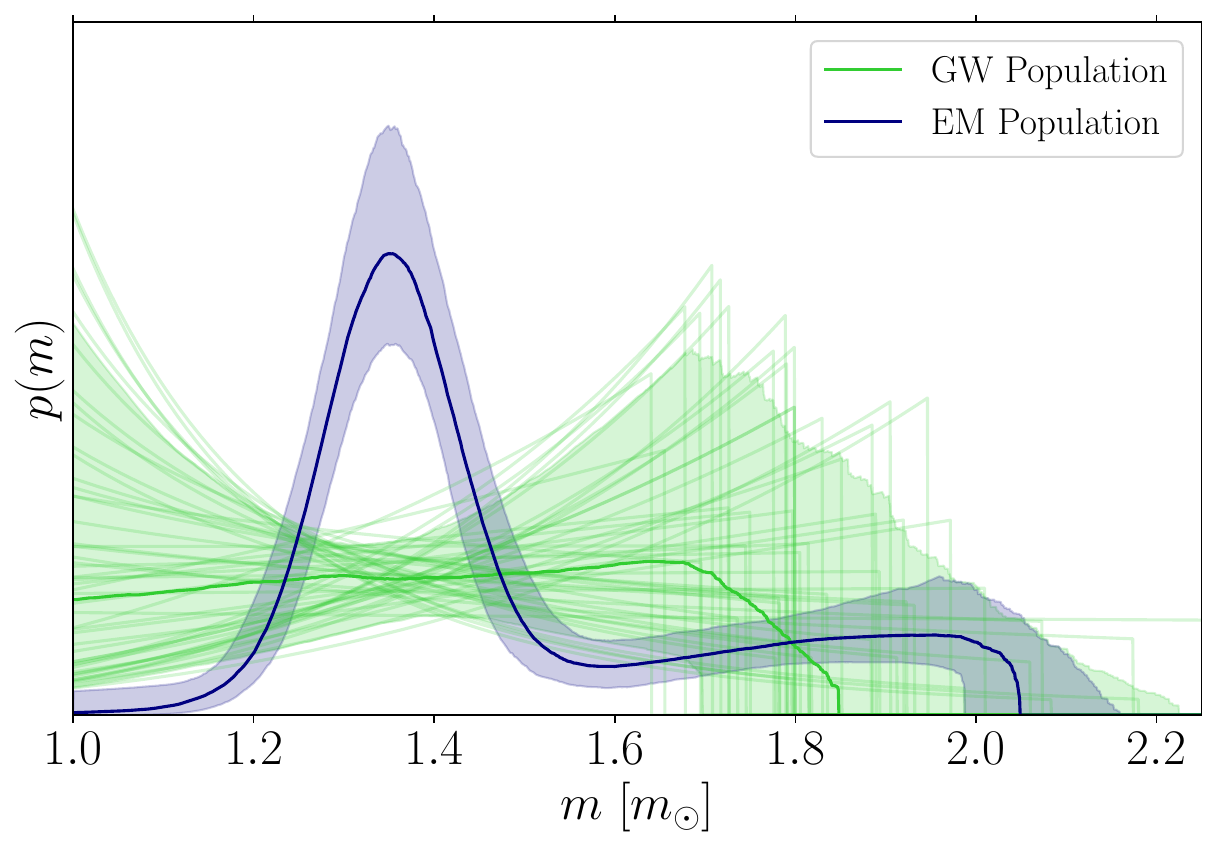}
    \caption{Posterior on the mass distribution of the GW BNS (orange) and the Galactic NS (blue) population. We plot the median and 90\%  highest-probability credible regions.
    The EM population is constrained to much better precision than the GW one due to the low number of GW BNS detections. 
    With the caveat that they correspond to the astrophysical BNS and observed Galactic NS distributions respectively, we find that the two distribution are inconsistent, in agreement with Ref.~\cite{KAGRA:2021duu}. Faint lines are random draws from the GW mass distribution, illustrating the bimodal uncertainties in the mass distribution.}
    \label{fig:mass_spec}
\end{figure}

\begin{figure}
    \centering
    \includegraphics[width=0.95\linewidth]{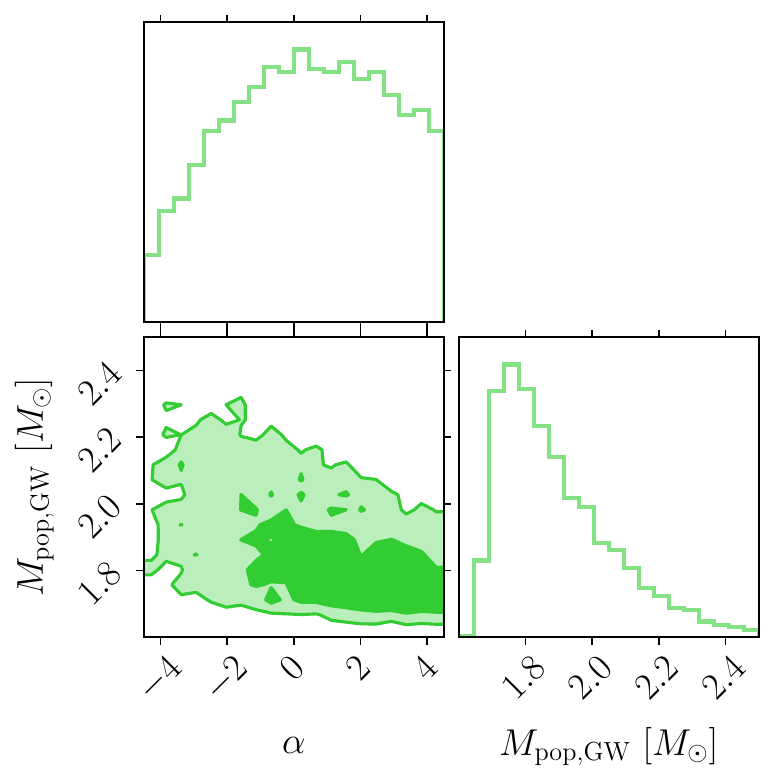}
    \caption{Marginalized posterior for the power-law slope $\alpha$ and maximum mass $M_{\rm pop,GW}$ of the GW population. The slope $\alpha$ is poorly constrained and thus its posterior rails against the upper prior bound, in turn affecting the $M_{\rm pop,GW}$ posterior. }
    \label{fig:alpha_mpop}
\end{figure}

Figure~\ref{fig:mass_spec} shows the inferred mass distribution of merging BNSs observed with GWs (modeled with a truncated power-law) and the observed distribution of Galactic NSs observed with EM (modeled with a truncated Gaussian mixture).
The BNS population is consistent with being flat and has large uncertainties due to the now number of events (a total of $4$ NSs).
The smallest uncertainty is at ${\sim}1.4\,M_{\odot}$, corresponding to the relatively well-measured masses on GW170817, while there is vanishing support for masses above ${\sim} 2.2\,M_\odot$ with $\mpopgw=\posteriormpopgw\,M_{\odot}$.
This shape is broadly consistent with the results of Refs.~\cite{KAGRA:2021duu, Landry:2021hvl} that additionally considered the two NSs in the NSBH binaries GW200105 and GW200115 and did not model the EoS.
The seemingly ``bimodal" shape with peaks at high and low masses at the 90\% level is model-dependent: it is an outcome of the fact that the distribution is well-measured at ${\sim}1.4\,M_{\odot}$ and we model it with a truncated power-law. 
Figure~\ref{fig:alpha_mpop} indeed shows that the power-law index $\alpha$ and the maximum mass, $\mpopgw$, are correlated and the upper limit on $\mpopgw$ depends on the $\alpha$ prior.  
In particular, while the one-dimensional posterior peaks at $\alpha\approx 0$, $\alpha \gtrsim 4$ cannot be ruled out but is only consistent with $\mpopgw \lesssim 2.0 M_{\odot}$.

The observed EM population is comparatively better constrained as it is based on a total of 74 pulsar mass measurements.
We find consistent results with Refs.~\cite{Alsing:2017bbc, FarrChatziioannou2020} that used the same pulsar mass data but did not infer the EoS with $\mu_1=\posterioremmuone\, M_{\odot}$  and $\mu_2=\posterioremmutwo\,M_{\odot}$, $f = \posterioremfracone$, and $\sigma_1=\posterioremsigmaone\,M_{\odot}$ and $\sigma_2=\posterioremsigmatwo\,M_{\odot}$.
The maximum mass is $\mpopem = \posteriormpopem\,M_{\odot}$, 
compared to $2.12^{+0.12}_{-0.17}\,M_{\odot}$ in~\cite{Alsing:2017bbc} and $2.25_{-0.26}^{+0.82}\,M_{\odot}$ in~\cite{ FarrChatziioannou2020}.
Our estimate is lower due to the fact that we simultaneously infer the EoS and impose $\mpopem < M_{\rm TOV}$.

Assuming that the three NICER pulsars are part of the general Galactic NS population leads to updated mass inference.
The original mass estimates quoted in Refs.~\cite{Miller:2019cac, Miller:2021qha, Choudhury:2024xbk} refer to flat mass priors, while our analysis effectively updates the prior to be the population distribution~\cite{FarrChatziioannou2020}.\footnote{The same is true for the two GW events, but the effect is minimal as the mass distribution uncertainty is wide and consistent with flat which was the inference prior to begin with.} 
The mass for each NICER target under a population-informed (flat) prior is $\popinformedjzerozeromass\,(\result{{1.44}^{+0.25}_{-0.23}})\,M_{\odot}$ for J0030+0451, $\popinformedjzerofourmass\, (\result{{1.42}^{+0.06}_{-0.06}})\,M_{\odot}$  for J0437-4715,  and $\popinformedjzerosevenmass\,(\result{{2.07}^{+0.11}_{-0.12}})\,M_{\odot}$ for J0740+6620. 
The J0740+6620 result is somewhat larger than the value in \citet{FarrChatziioannou2020},  $ 2.03^{+0.17}_{-0.14}\,M_{\odot}$.
The effect is most stark for J0030+0451 whose mass is poorly measured from the X-ray data alone, but now resides in the dominant peak of the mass distribution.

\subsection{Constraints on EoS quantities}
\label{sec:eos}

\begin{figure*}
    \centering
    \includegraphics[width=0.99\textwidth]{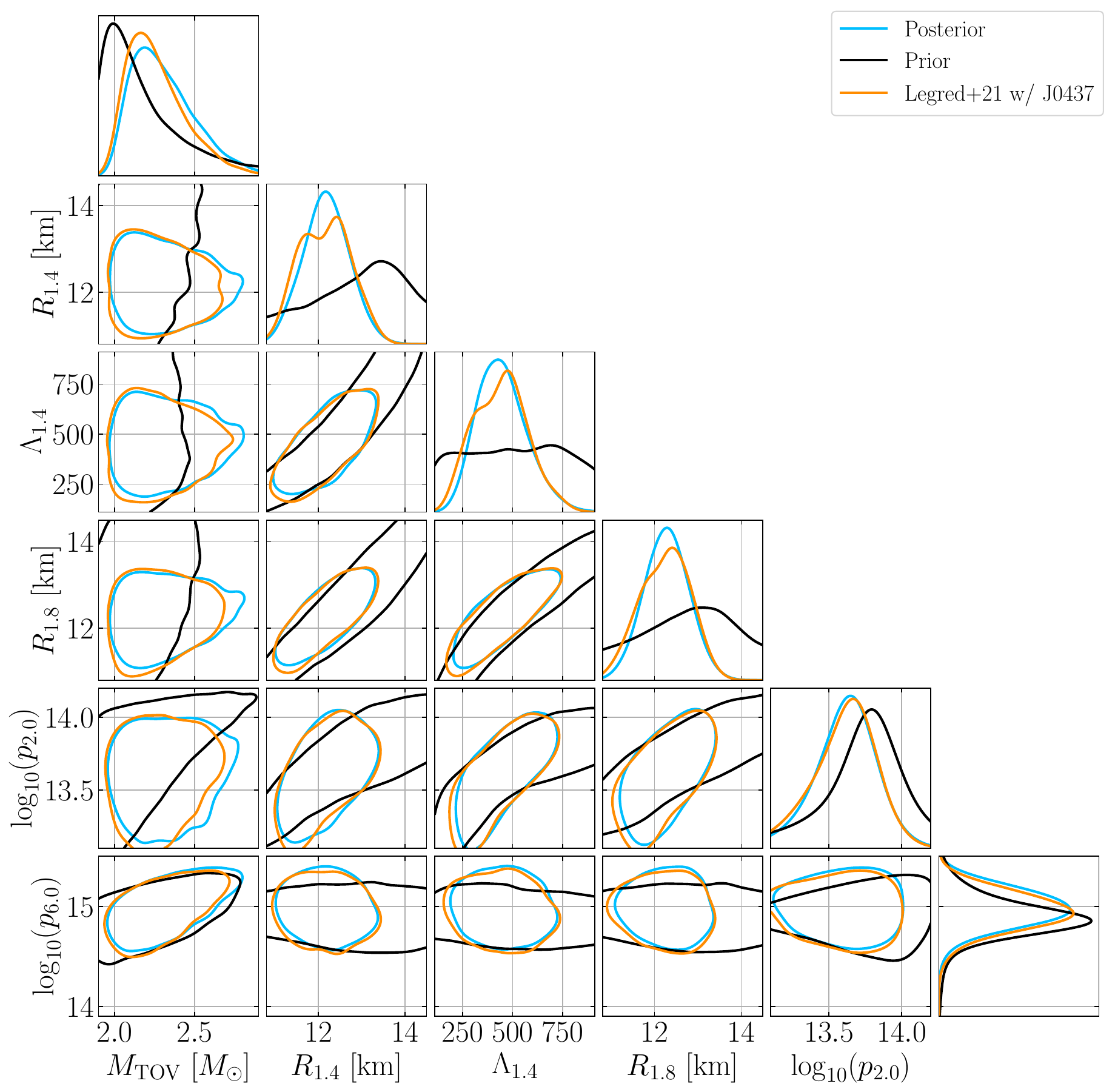}
    \caption{One- and two-dimensional posteriors for select EoS macroscopic and microscopic parameters: the TOV mass, $M_{\rm TOV}$, the radius and tidal deformability of a canonical $1.4\,M_{\odot}$ NS, $R_{1.4}$ and $\Lambda_{1.4}$ respectively, the radius of a $1.8\,M_{\odot}$ NS, $R_{1.8}$, and the log-base-10 pressure (divided by the speed of light squared) at twice and 6 times nuclear saturation, $p_{2.0}$ and $p_{6.0}$ respectively, when measured in $\rm{g}/\rm{cm}^3$.
    Two-dimensional contours denote the boundaries of the 90\% credible regions.
    We show the prior (black), the posterior from the main analysis that marginalizes over the mass distribution (blue), and the analogous posterior that arises from additionally including the mass-radius measurement of J0437-4715 in the analysis of Ref.~\cite{Legred:2021}. }
    \label{fig:eos-properties-corner}
\end{figure*}

\begin{figure}
    \centering
    \includegraphics[width=0.5\textwidth]{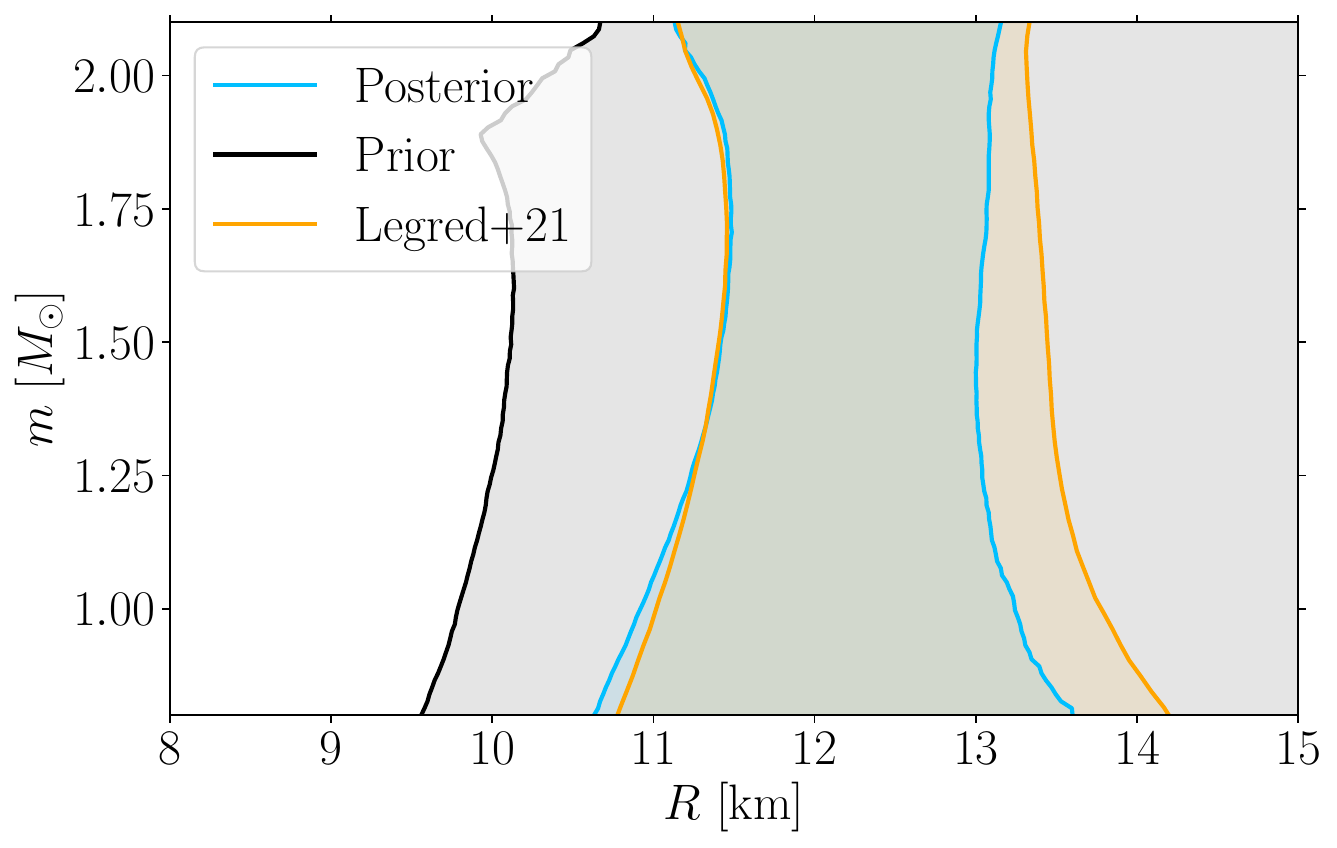}
    \caption{Mass-radius inference, we show the $90 \%$ symmetric credible region for the radius at each mass.  
    We plot the prior (black), posterior from the main analysis that marginalizes over the mass distribution (blue), and posterior from Ref.~\cite{Legred:2021} that fixes the mass distribution to flat and does not include J0437-4715.
    The upper limit on the radius decreases by $\sim 0.5\,\rm{km}$ for all masses.
    }
    \label{fig:r-of-m-quantiles}
\end{figure}

\begin{figure}
    \centering
    \includegraphics[width=0.5\textwidth]{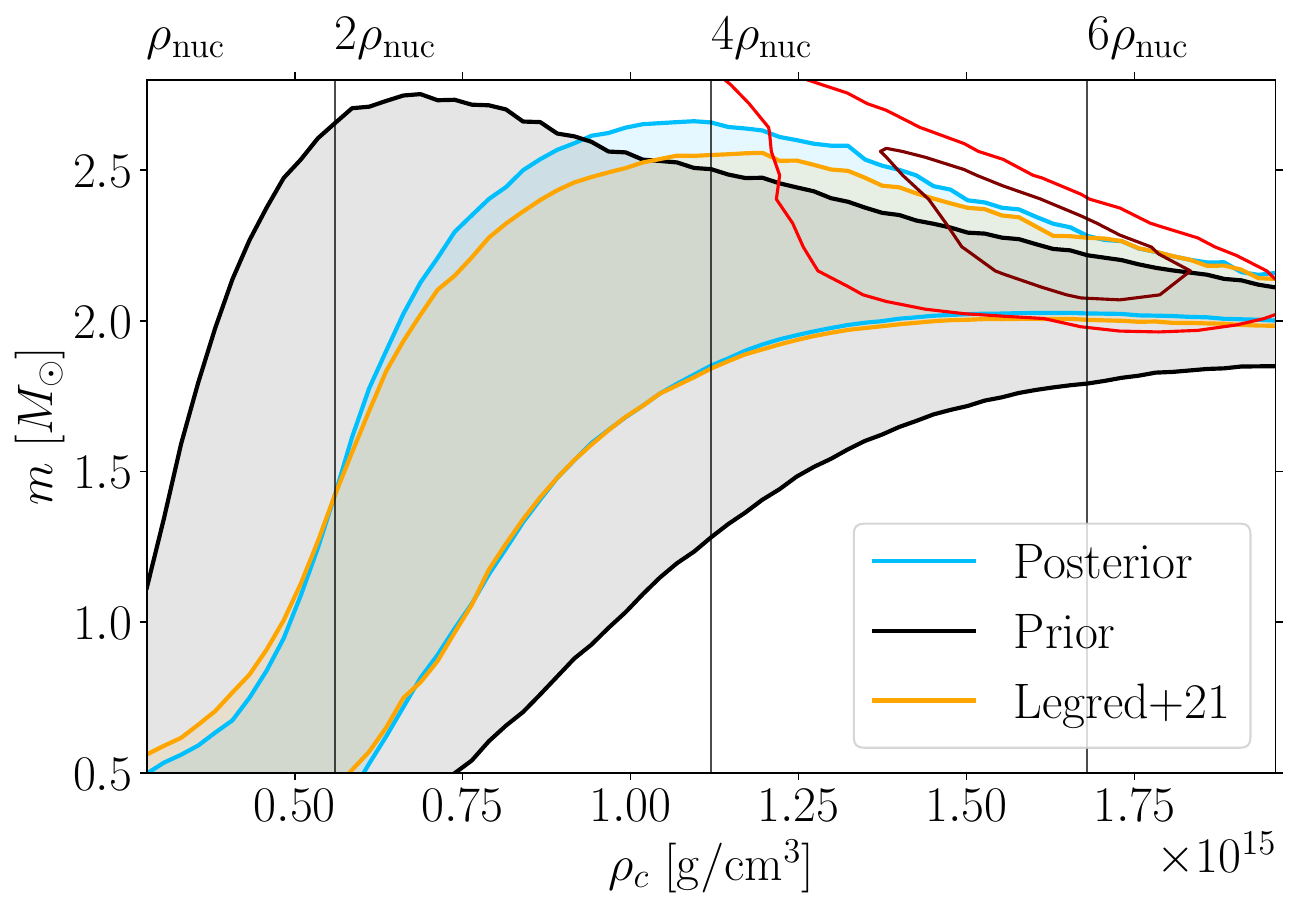}
    \caption{Mass-central density inference, we show the $90 \%$ symmetric credible region for the NS mass at each value of the central density $\rho_c$.  
    We plot the prior (black), posterior from the main analysis that marginalizes over the mass distribution (blue), and posterior from Ref.~\cite{Legred:2021} that fixes the mass distribution to flat and does not include J0437-4715.
    Vertical lines denote multiples of the nuclear saturation density.  
    Maroon and red contours mark $1$ and $2$-$\sigma$ credible regions, respectively,  for the joint posterior on $\rho_{c}$-$M_{\rm TOV}$.
    }
    \label{fig:m-of-rhoc-quantiles}
\end{figure}

\begin{figure}
    \centering
    \includegraphics[width=.5\textwidth]{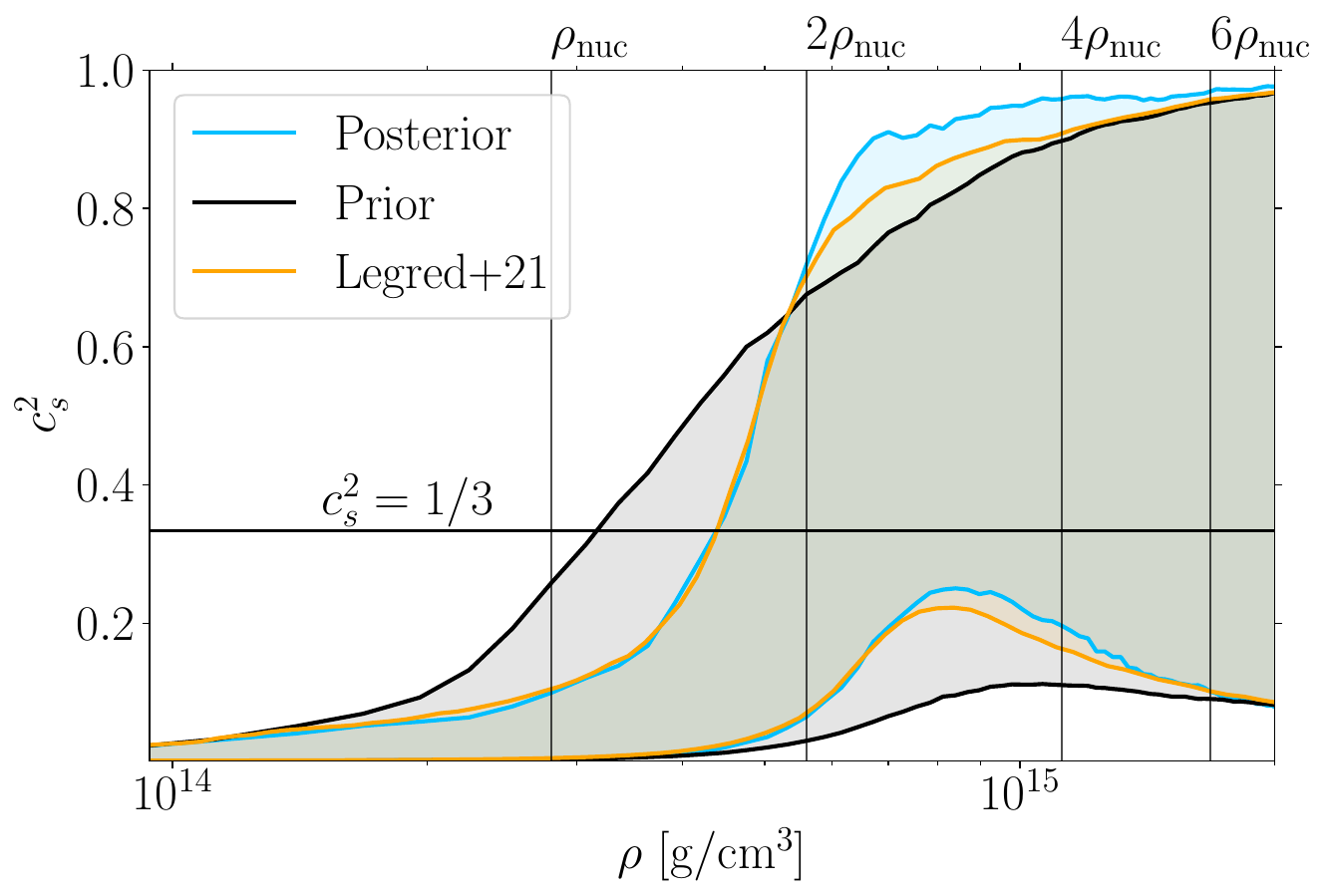}
    \caption{Speed of sound-density inference, we show the $90 \%$ symmetric credible region for the speed of sound squared, $c_s^2$ at each rest-mass density $\rho$.  
    We plot the prior (black), posterior from the main analysis that marginalizes over the mass distribution (blue), and posterior from Ref.~\cite{Legred:2021} that fixes the mass distribution to flat and does not include J0437-4715.
    Vertical lines denote multiples of the nuclear saturation density. The speed of sound increases by $\sim 5\%$ around densities $2-3$ times saturation density. }
    \label{fig:cs2-quantiles}
\end{figure}

\begin{figure}
    \centering
    \includegraphics[width=.5\textwidth]{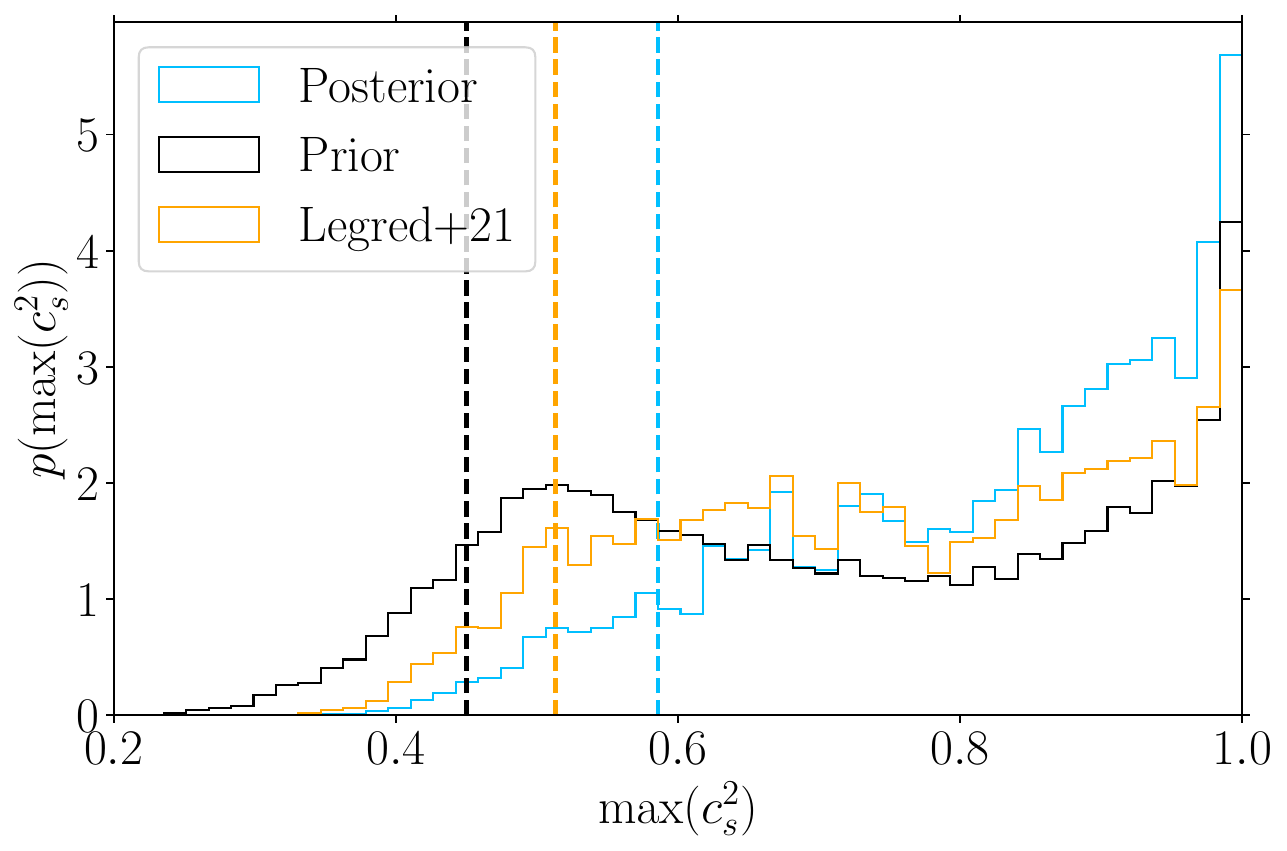}
    \caption{Marginalized posterior for the maximum speed of sound squared inside a stable NS.  We plot the prior (black), posterior from the main analysis that marginalizes over the mass distribution (blue), and posterior from Ref.~\cite{Legred:2021} that fixes the mass distribution to flat and does not include J0437-4715.
    The $90\%$ lower limit on the maximum speed of sound, marked by dashed vertical lines, increases from ${\sim} \legredcsmaxlowerbound$ to ${\sim} \posteriorcsmaxlowerbound$.}
    \label{fig:max-cs2}
\end{figure}

Figure~\ref{fig:eos-properties-corner} shows the prior and posterior for various macroscopic and microscopic EoS properties: the TOV mass, $M_{\rm TOV}$, the radius and tidal deformability of a canonical $1.4\,M_{\odot}$ NS, $R_{1.4}$ and $\Lambda_{1.4}$ respectively, the radius of a $1.8\,M_{\odot}$ NS, $\Lambda_{1.8}$, and the pressure at twice and 6 times nuclear saturation, $p_{2.0}$ and $p_{6.0}$ respectively.
We infer $\Lambda_{1.4} = \posteriorLambdatyp$ and $R_{1.4} = \posteriorRtyp\,$km.  
For comparison, we also plot the corresponding analysis from \citet{Legred:2021} that fixes all mass distributions to uniform.
To isolate the effect of the mass distribution inference, we repeat the analysis of Ref.~\cite{Legred:2021} while adding the X-ray mass-radius measurement of J0437-4715 such that the two analyses use the same NICER and GW data.
We obtain largely consistent results: mass-marginalization leads to mild changes in $R_{1.4}$ and $\Lambda_{1.4}$, while including spider pulsars in the analysis and introducing an EoS-limited astrophysical maximum mass leads to a mild increase in the inferred value of $M_{\rm TOV}$.

These results are consistent with previous estimates.
\citet{Legred:2021} used the GP EoS model with the same GW dataset, the first two NICER objects, J0030+0451 and J0740+6620, and the mass of J0348+0432 (all with a fixed flat mass prior) to find $R_{1.4} = 12.6^{+1.0}_{-1.1}\,$km and $M_{\rm TOV}=2.21^{+0.31}_{-0.21}\,M_\odot$.
Our updated radius estimate has a ${\sim} 0.4 \,\rm{km}$ lower median due to the new J0437-4715 data that favor softer EoSs and a ${\sim} 20$\% smaller uncertainty due to the fact that we use more NICER and massive pulsar data.
Our updated $M_{\rm TOV}$ estimate of $\posteriorMtov\,M_\odot$ is marginally larger than the value found in \citet{Legred:2021}, which can be attributed to the spider pulsars, and the removal of the EoS Occam penalty for massive pulsar measurements, see the Appendix of Ref.~\cite{Legred:2021}. 

The full mass-radius inferred relation is shown in Fig.~\ref{fig:r-of-m-quantiles} 
which plots the $90\%$ symmetric credible region for the radius at each mass.
We include the prior, the posterior from our analysis, and compare against the posterior from \citet{Legred:2021}, i.e., without J0437-4715. 
While the radius lower limit is broadly consistent with Ref.~\cite{Legred:2021}, we obtain a lower radius upper limit for all masses by ${\sim}500$\,m, which we attribute to the new data for the J0437-4715 radius. 
We additionally plot credible regions for the relation between the NS mass $m$ and its central density $\rho_c$ in Fig.~\ref{fig:m-of-rhoc-quantiles}.
The upper limit on the mass of a NS with central density $4$ times the nuclear saturation density ($\rho_{\nuc}$) increases from ${\sim} \legredmatfourrhonucupperbound\,M_{\odot}$ to ${\sim} \posteriormatfourrhonucupperbound\, M_{\odot}$, primarily due to the removal of the Occam penalty and the inclusion of spider pulsars.  
The central density of the maximum mass star is inferred to be $\posteriorrhocatmmax~\rho_{\nuc}$ (red contours).

We examine the EoS microscopic properties and specifically the speed of sound as a function of density in Fig.~\ref{fig:cs2-quantiles} and the maximum speed of sound inside NSs in Fig.~\ref{fig:max-cs2}.

Compared to \citet{Legred:2021}, our analysis favors a larger speed of sound around $2-4 \rho_{\nuc}$ and a larger maximum speed of sound throughout.
 The $90\%$ lower limit on the maximum speed of sound, increases from $\sim \legredcsmaxlowerbound$ in Ref.~\cite{Legred:2021} to $\sim \posteriorcsmaxlowerbound$ for our analysis.
This higher maximum speed of sound is necessary to explain the high mass of certain Galactic pulsars which, though poorly measured, can have exceptionally large median values, e.g., J01748-2021B with an estimated mass of $2.74^{+0.21}_{-0.21}\,M_\odot$~\citep{Freire:2007jd} at 68\% credibility.
The addition of the NICER radius measurement J0437-4715 also marginally impacts the inferred maximum sound speed; removing the radius measurement of J0437-4715, Appendix~\ref{sec:no-nicer}, leads to a maximum $c_s^2$ value of $\nojzerofourcsmax$.

\subsection{Joint constraints on the population and EoS}
\label{sec:joint}

\begin{figure}
    \centering
    \includegraphics[width=.48\textwidth]{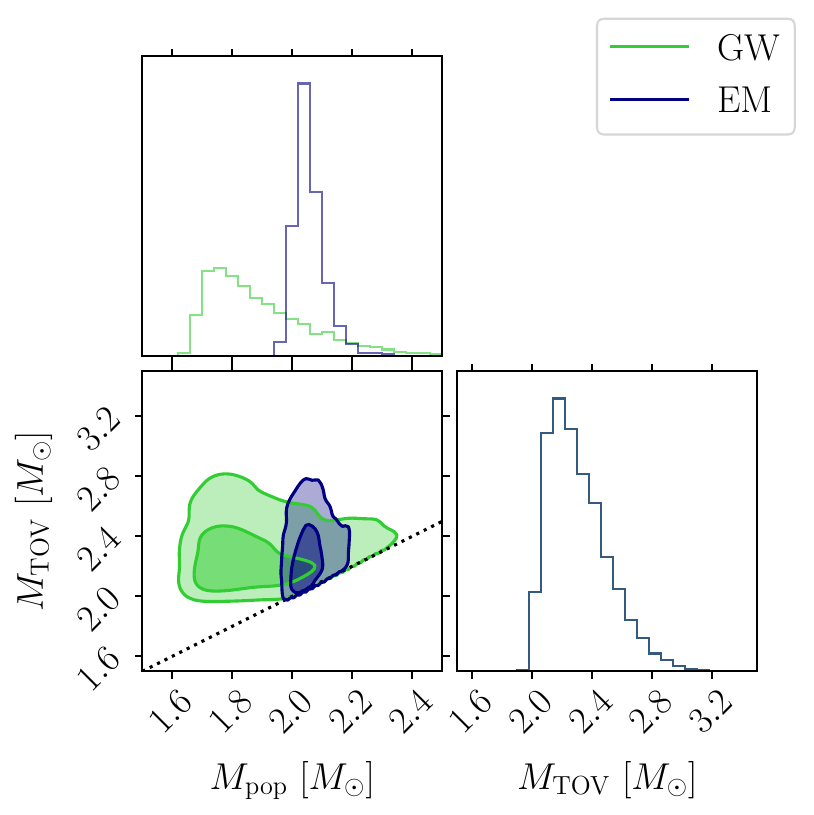}
    \caption{One-and two-dimensional posteriors for $M_{\rm TOV}$ and the maximum astrophysical mass $M_{\pop}$ for the Galactic NSs (blue) and the merging BNSs (orange). 
    The black dashed line represents $M_{\rm pop} = M_{\rm TOV}$, which is imposed in our analyses as we assume that all objects are NSs. 
    The TOV mass is consistent with the astrophysical maximum mass for both populations. Contours are drawn at 50\% and 90\% levels.
    }
    \label{fig:mpopvsmtov}
\end{figure}

The joint EoS-mass inference allows us to separate the TOV mass, $M_{\rm TOV}$, from the maximum astrophysical mass in the two subpopulations, $\mpopem$ and $\mpopgw$.
Figure~\ref{fig:mpopvsmtov} shows the joint posterior for $M_{\rm TOV}$ and the two population maximum masses, denoted collectively as $M_{\pop}$.
The limit $M_{\rm TOV} = M_{\pop}$ is marked with a dashed line; points near the line correspond to maximum population masses that are equal to the TOV mass.  
As also evident in Fig.~\ref{fig:mass_spec}, the two population maximum masses are consistent with each other within their statistical uncertainties.
The difference between the maximum mass in the EM (GW) population and $M_{\rm TOV}$ is less than $\mpopemmtovdiffninety$ ($\mpopgwmtovdiffninety$) at 90\% credibility.

We therefore have no evidence that the maximum mass of neutron stars formed astrophysically is different than the maximum mass possible from nuclear physics.

\section{Conclusions}
\label{sec:conclusions}

As a first step toward untangling the properties of NSs that depend on nuclear physics versus astrophysics, in this study we presented a joint inference of the dense matter EoS and the NS mass distribution.
We considered two subpopulations of NSs corresponding to merging BNSs observed with GWs and Galactic NSs observed with EM.
All NSs share the same universal EoS modeled with a flexible GP mixture.
Our results are consistent with existing EoS-only or mass-only inference where applicable~\citep{Legred:2021, FarrChatziioannou2020, KAGRA:2021duu}.
However, the joint inference scheme allows us to begin addressing the interplay between nuclear physics and astrophysics in determining NS observational properties. 
Focusing on NS masses, we find no evidence that the maximum mass of NSs observed with either EM or GWs is different than the maximum mass allowed by nuclear physics.
Moreover, we updated the estimates of the canonical NS radius and the TOV mass to $R_{1.4}=\posteriorRtyp\,\rm{km}$ and $M_{\rm TOV}=\posteriorMtov M_{\odot}$, respectively.

\subsection{Past work}

Our results are broadly consistent with comparable studies. 
Whereas we model the EoS phenomenologically as a GP, \citet{Rutherford:2024srk} used a piecewise-polytropic EoS model and the same data as \citet{Legred:2021} plus the radius measurement of J0437-4715; they found $R_{1.4}=12.3^{+0.5}_{-0.8}\,$km.
Our result has a ${\sim} 30\%$ larger uncertainty likely due to the more flexible EoS model. 

\citet{Fan:2023spm} simultaneously inferred the mass distribution and the EoS, though they assumed the same mass distribution for all NSs, and that the upper truncation mass for the NS population is $M_{\rm TOV}$. 
They used the same data as our study except the radius measurement of J0437-4715, and included ${\sim} 50$ additional pulsar mass measurements. 
They used a variety of parameteric and nonparametric EoS models, but recovered similar values of $R_{1.4}$ and $M_{\rm TOV}$ for all models, indicating their nonparametric models may have limited flexibility (analogous to the ``model-informed prior" of \cite{Landry:2018prl, Essick:2019ldf}). 
They further incorporated information from perturbative quantum chromodynamics (pQCD) at high densities, and chiral perturbation theory at low densities, both of which strongly informed the estimate of $M_{\rm TOV}$ due to the choice of modeling of correlations.  
They found $M_{\rm TOV} = 2.25^{+0.08}_{-0.07}\,M_{\odot}$. 

\citet{Biswas:2024hja} also simultaneously inferred the population and the EoS, similarly requiring the NSs to form a single population which is truncated by $M_{\rm TOV}$.
For the EoS they used a piecewise-polytropic parameterization, hybridized with a low-density prescription constrained by chiral effective field theory.
They analyzed the same data as \citet{Fan:2023spm}, and additionally the PREX-II~\cite{Adhikari:2021phr} and CREX~\cite{CREX:2022kgg} measurements of the neutron skin thickness of $^{208}\rm{Pb}$ and $^{48}\rm{Ca}$ respectively.

They found $R_{1.4}=12.5^{+0.3}_{-0.3}\,$km, and $M_{\rm TOV} = 2.27^{+0.08}_{-0.09}\,M_{\odot}$.  
These uncertainties are substantially lower than our results. 
The radius constraint can at least in part be attributed to information from chiral perturbation theory, while the EoS parameterization also results in tighter inference throughout due to less modeling flexibility~\cite{Essick:2020flb, Legred:2022pyp}.
Moreover, the use of a single mass distribution places a very strong prior on the masses of the GW events, with the mass of GW170817 for example likely tightly constrained to be within the primary peak of the bimodal mass distribution.
Such improved mass measurement will translate to tighter tidal and hence EoS constraints.
The impact of pQCD information~\cite{Komoltsev:2021jzg} remains unclear~\cite{Somasundaram:2022ztm, Komoltsev:2023zor}, though the prescription used in that analysis is likely informative of $M_{\rm  TOV}$.

Other studies have obtained multimessenger constraints on the EOS by combining GW, gamma-ray burst, and kilonova observations surrounding GW170817 with fits to the EM emission from BNS simulations \citep{Radice:2017lry, Coughlin:2018miv, Coughlin:2018fis, Pang:2022rzc}. While there are systematic and statistical uncertainties in the models and observations, these studies infer $R_{1.4}$ and $\Lambda_{1.4}$ broadly consistent with our results.

\subsection{Caveats}

Our findings depend on several analysis choices and assumptions. 
In the appendices, we examine their impact, and here we summarize our conclusions.

In our main analysis, we assume that selection biases in the radio and X-ray surveys are negligible. 
In Appendix~\ref{app:uniform_em_population} we consider the impact that modeling all Galactic NSs with the same bimodal distribution without taking selection effects into account has.
Compared to an analysis that fixes the pulsar mass distribution to uniform up to $M_{\rm TOV}$~\cite{Legred:2021}, inference of the mass distribution leads to an EoS that is marginally softer at low densities and marginally stiffer at high densities. 
As a consequence, the evidence for a violation of the conformal limit $c_s^2=1/3$ increases and the lower limit on the maximum speed of sound increases by ${\sim} 10\%$.

Data selection further influences our results. In particular, different interpretations of the NICER observations exist in the literature. 
Given systematic studies on the impact of analysis assumptions on NICER measurements~\cite{Essick:2021ezv, Vinciguerra:2023qxq} we present results without J0030+0451 and/or J0437-4715 in  Appendix~\ref{sec:no-nicer}. 
Excluding J0437-4715 leads to a stiffer inferred EoS with $R_{1.4} = \nojzerofourRtyp\, \rm{km}$ and consistent results with Ref.~\cite{Legred:2021}.  
Excluding J0030+0451 results in a substantially reduced value of $R_{1.4} = \nojzerozeroRtyp\, \rm{km}$.
However, all results are consistent with each other at $90\%$ credibility, see Fig.~\ref{fig:R14vsnoNICER} in Appendix~\ref{sec:no-nicer}. 

Additionally, our main results assume a fixed spin distribution, extending in magnitude up to 0.05 for GW170817 and 0.4 for GW190425. 
Assumptions about the spin affect mass inference through the mass-spin correlation~\cite{Cutler:1994ys} and hence mass population inference.
We explore the impact of restricting the spin of GW190425 further in Appendix~\ref{app:lowspin}.
Imposing an upper limit of 0.05 results in a tighter constraint on its mass ratio and a lower primary mass, which correspondingly reduces the value of $M_{\rm pop,GW}$.
Consistency between $\mpopgw$ and $M_{\rm TOV}$ is reduced with their difference less than $\mpopgwmtovdiffninetylowspin$ at 90\% credibility.
Therefore we still find no strong evidence that the TOV and the maximum astrophysical mass are different. Simultaneous inference of the spin distribution~\cite{Biscoveanu:2021eht}, along with the EoS and mass distribution, is reserved for future work.

Finally, in this study, we restricted to two subpopulations of NSs: GW observations of BNSs and Galactic NSs from radio or X-ray surveys. 
As a consequence, our mass distribution inference is only predictive below $2.5\,M_{\odot}$, which we took to be the (fixed) demarcation between NSs and BHs. Extending to higher masses would require simultaneously classifying GW events as BNSs, NSBHs or BBHs within the analysis framework~\cite{Essick:2020ghc,Chatziioannou:2020msi}, while introducing a third NS subpopulation associated with the NSBH mergers. 
This would allow us to treat other GW discoveries, such as $\rm GW230529\_181500$~\cite{LIGOScientific:2024elc} and GW190814~\cite{LIGOScientific:2020zkf}, whose nature is ambiguous. 
These and further extensions to the joint inference methodology presented here will become necessary to fully explore the interplay between nuclear physics and astrophysics on the properties of NSs as our catalog of informative NS observations increases in size.

\acknowledgments

We thank Will Farr for helpful discussions on hierarchical inference of subpopulations. We are also grateful to Reed Essick for useful discussions on population modeling with our dataset. We also thank Sylvia Biscoveanu for helpful comments on the manuscript.
J.G. would like to gratefully acknowledge the support from the National Science Foundation through the Grant NSF PHY-2207758.
I.L. and K.C. acknowledge support from the Department of Energy under award number DE-SC0023101, the Sloan Foundation, and by a grant from the Simons Foundation (MP-SCMPS-00001470).
P.L. is supported by the Natural Sciences \& Engineering Research Council of Canada (NSERC). Research at Perimeter Institute is supported in part by the Government of Canada through the Department of Innovation, Science and Economic Development and by the Province of Ontario through the Ministry of Colleges and Universities.
The authors are grateful for computational resources provided by the LIGO Laboratory and supported by National Science Foundation Grants PHY-0757058 and PHY-0823459.
This material is based upon work supported by NSF's LIGO Laboratory which is a major facility fully funded by the National Science Foundation.
Software: \texttt{bilby}~\cite{Ashton2019, Romero-Shaw:2020owr}, \texttt{dynesty}~\cite{Speagle20}, \texttt{scipy}~\cite{scipy}, \texttt{numpy}~\cite{numpy}, \texttt{matplotlib}~\cite{matplotlib}, \texttt{lwp}~\cite{lwp}.

\appendix

\section{Reweighting scheme for the joint posterior}\label{app:likelihood}

The joint posterior for the GP EoS $\eos$ and the population hyperparameters $\eta=\{\eta_{\GW},\eta_{\EM}\}$ is~\cite{Landry:2020vaw,Chatziioannou:2020pqz}
\begin{equation}
\label{posterior-full}
    p(\eos, \eta|d) = \frac{\mathcal{L}(d|\eos, \eta) \pi(\eos, \eta)}{p(d)}\,,
\end{equation}
where $d$ is the data, $\mathcal{L}(d|\eos, \eta)$ is the likelihood, $\pi(\eos, \eta)$ is the prior, and $p(d)$ is the evidence.  
We choose a prior of $\pi(\eos, \eta) = \pi(\eos) \pi(\eta) \Theta(M_{\rm TOV}- \mpopem)\Theta(M_{\rm TOV}- \mpopgw)$, where $\pi(\epsilon)$, is the model agnostic prior defined in Refs.~\cite{Landry:2018prl, Essick:2019ldf} (uniform over GP draws), and $\pi(\eta)$ is the prior on the population hyperparameters, as described in the main text (uniform over all parameters). 
Since the GW and EM datasets are independent, the total likelihood factors into individual likelihoods
\begin{equation}
    \mathcal{L}(d|\eos, \eta)
        = \mathcal{L}_{\GW}(d|\eos, \eta_{\GW}) \mathcal{L}_{\rm NICER}(d|\eos, \eta_{\EM})\nonumber  \mathcal{L}_{\rm PSR}(d|\eos, \eta_{\EM})\,,
\end{equation}
given in Eqs.~\eqref{eq:GWlikelihood},~\eqref{eq:NICERlikelihood}, and~\eqref{eq:psrlikelihood} respectively. 

We evaluate the likelihood $\mathcal{L}(d|\eos, \eta)$ with a reweighting scheme based on a simpler lower-dimensional EoS model $\eos_0$, details about which are given in Appendix~\ref{app:EoS-simple}. We first obtain samples from the joint posterior for $\eos_0$ and $\eta$ using standard stochastic sampling~\cite{Ashton2019}.
\begin{equation}
    p_0(\eos_0, \eta|d) = \frac{\mathcal{L}_0(d|\eos_0, \eta) \pi_0(\eos_0, \eta)}{p_{0}(d)}\,.
\end{equation}
We then use the marginal mass distribution posterior
\begin{equation}
\label{eq:proposal-marginal}
    p_0(\eta|d)=\int p_0(\eos_0, \eta|d)d\eos_0\,,
\end{equation}
as a proposal distribution to rewrite Eq.~\eqref{posterior-full} as
\begin{equation}
    \label{eq:reweighting}
    p(\eos, \eta|d)  \propto\mathcal{L}(d|\eos, \eta) \frac{\pi(\eta)\Theta(M_{\rm TOV}- M_{\pop})}{p_0(\eta|d)} p_0(\eta|d)\pi(\eos)\,,
\end{equation}
where we have dropped the normalization $p(d)$ and defined $\Theta(M_{\rm TOV}- M_{\pop})\equiv\Theta(M_{\rm TOV}- M_{\pop, EM})\Theta(M_{\rm TOV}- M_{\pop, GW})$.
Reweighting includes
\begin{enumerate}
\item Compute a Kernel Density Estimate (KDE) of $p_0(\eta|d)$ so that we can directly evaluate the density for each value of $\eta$.
\item Draw samples $\eos\sim\pi(\eos)$ and $\eta\sim p_0(\eta|d)$. If $M_{\rm TOV}< \mpopem$ or $M_{\rm TOV}< \mpopgw$, reject the sample.
\item For accepted $(\eos,\eta)$ samples compute the weight 
\begin{equation}
w=\mathcal{L}(d|\eos, \eta)\frac{\pi(\eta) }{p_0(\eta|d)}  \,.
\end{equation} 
The term $p_0(\eta|d)$ is computed with the KDE from step \#1 and 
the likelihood $\mathcal{L}(d|\eos, \eta)$ is computed with a Monte Carlo sum over individual-event posterior samples.
\item Each sample $(\eos,\eta)$ is a weighted draw from the joint posterior $p(\eos,\eta|d)$ with weight $w$.
\end{enumerate}

In practice, we consider the EM likelihood for the two EM datasets

\begin{equation}
    \mathcal{L}(d_{\EM}| \eos, \eta_{\EM}) = \mathcal{L}(d_{\NICER}| \eos, \eta_{\EM}) \times \mathcal{L}(d_{\PSR}| \eos, \eta_{\EM})\,,
\end{equation}
and the combined likelihood

\begin{equation}
    \mathcal{L}(d|\eos, \eta) = \mathcal{L}(d_{\EM}|\eos, \eta_{\EM}) \times \mathcal{L}(d_{\GW}|\eos, \etagw)\,,
\end{equation}
from Eq.~\eqref{eq:reweighting}.
In order to calculate the likelihood for the GW population parameters $\eta_{\GW}$, we approximate

\begin{equation}
\begin{split}
    & \mathcal{L}(d|\eta_{\GW}) =\\& \int \mathcal{L}(d_{\EM}|\eta_{\EM}, \eos) \mathcal{L}(d_{\GW}|\eta_{\GW}, \eos) \pi(\eta_{\EM}, \eos) d\eos \, d\eta_{\EM} 
\end{split}
\end{equation}

with the Monte Carlo sum:

\begin{equation}\label{eq:marginal_MC_mpopGW}
\begin{split}
    \mathcal{L}(d|\eta_{\GW}) \approx  & \sum_{\eos \sim \pi(\eos)} \mathcal{L}(d_{\GW}|\eta_{\GW}, \eos) \times \\
     \Big[ & \sum_{\eta_{\EM} \sim p_0(\eta_{\EM})} \frac{\mathcal{L}(d_{\EM}|\eta_{\EM}, \eos)}{p_0(\eta_{\EM}|d)} \pi(\eta_{\EM}|\eos) \Big]\,. 
\end{split}
\end{equation}
The likelihood for the EM population parameters is obtained by by swapping $\GW \leftrightarrow \EM$ in Eq.~\eqref{eq:marginal_MC_mpopGW}.

Similarly, we compute the likelihood for the EoS $\eos$  as

\begin{equation}\label{eq:marginal_MC_eos}
\begin{split}
    \mathcal{L}(d|\eos) \approx  & \sum_{\eta_{\GW} \sim p_0(\eta_{\GW})} \frac{\mathcal{L}(d_{\GW}|\eta_{\GW}, \eos)}{p_0(\eta_{\GW}|d)} \pi(\eta_{\GW}|\eos) \times \\
     & \sum_{\eta_{\EM} \sim p_0(\eta_{\EM})} \frac{\mathcal{L}(d_{\EM}|\eta_{\EM}, \eos)}{p_0(\eta_{\EM}|d)} \pi(\eta_{\EM}|\eos)\,. 
\end{split}
\end{equation}

\section{Approximate lower-dimensional EoS model}
\label{app:EoS-simple}

The reweighting scheme of Appendix~\ref{app:likelihood} utilizes a lower-dimensional EoS model $\eos_0$ that gets marginalized away in Eq.~\eqref{eq:proposal-marginal}, solely for constructing an efficient proposal distribution for the hyperparameters $\eta$.
The goal of including $\eos_0$ in the first place is to avoid potential systematic biases in $p_0(\eta|d)$ if inferred without any reference to an EoS~\cite{Golomb:2021tll}.
Such biases would make it an ineffective proposal distribution for the reweighting of Eq.~\eqref{eq:reweighting}.
Our requirement for $\eos_0$ is therefore that it can be evaluated efficiently and that it roughly captures typical EoS behaviors.
Existing parametric models such as the piecewise-polytropic~\cite{Read:2008iy}, spectral~\cite{Lindblom:2010bb}, or speed-of-sound~\cite{Tews:2018kmu, Greif:2020pju} models could play this role.
However, we find that something even simpler suffices.

We take advantage of the simple relation between the NS moment-of-inertia $I$ and mass $m$~\citep{Yagi:2013awa, Essick:2023fso} for hadronic EoSs. 
For EoSs without rapid changes in the speed of sound~\cite{Essick:2023fso},
\begin{equation}
    \frac{d\ln I}{d\ln m} \sim 1.6 \pm \mathcal{O}(10^{-2})\,.
\end{equation}
We therefore define $\eos_0$ with a linear relationship between $\ln I$ and $\ln m$:
\begin{equation}
    \ln I = a \ln m + b \,,
\end{equation}
where the free parameters $a$ and $b$ define a specific EoS. 
From the $I(m)$ relation we can obtain $\Lambda(m)$ (used for analyzing GW data) and $R(m)$ (used for analyzing X-ray data) with the $I$-Love~\cite{Yagi:2013bca} and $C$-Love~\cite{Yagi:2016bkt, Carson:2018xri} universal relations respectively.
Since the model does not have a miscrophysics interpretation, it does not self-consistently lead to a maximum-mass solution.
Instead we define its TOV mass as $\Lambda(M_{\rm TOV}) = \Lambda_{\rm thresh}=\exp(1.89)$ which empirically produces reasonable values for $M_{\rm TOV}$,
 
We find that this model is inexpensive to sample and accurate enough that that it leads to an improved reweighting efficiency.
However, it would not be a reliable model for EoS inference due to its simplistic nature.

\section{Method validation}
\label{app:validation}

\begin{figure}
    \centering
    \includegraphics[width=0.5\textwidth]{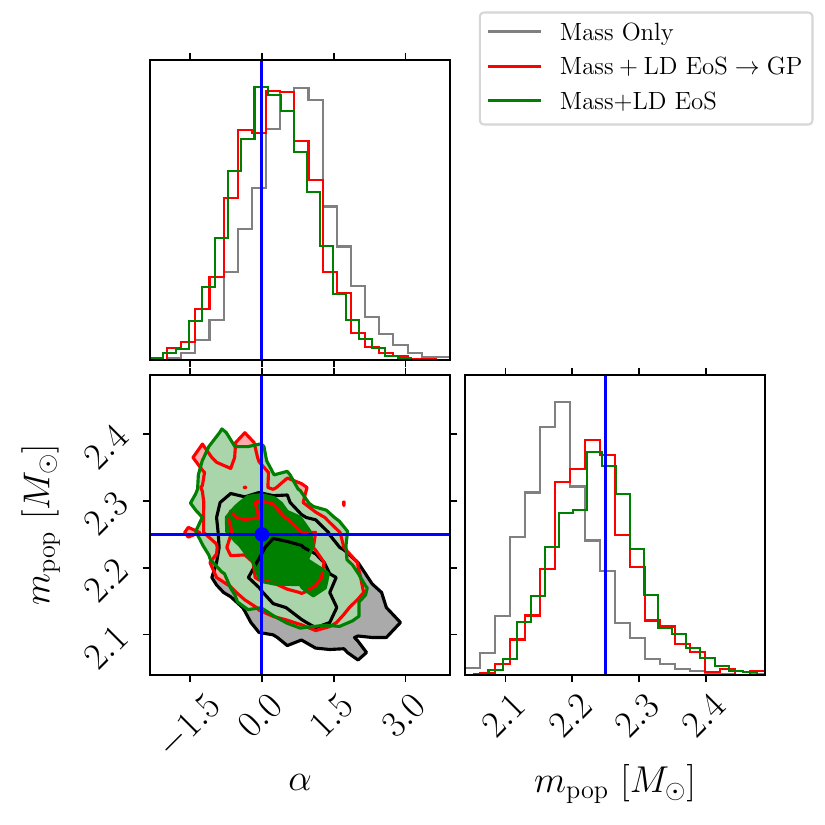}
    \caption{One- and two-dimensional posteriors for the mass distribution slope and maximum mass from 23 simulated BNSs.
    We plot mass-only population inference (grey) which defaults to the individual-event-inference prior on the tidal deformability, joint mass-EoS inference using the lower-dimensional EoS model (green) and the full mass-EoS joint inference with the GP EoS model (red).
    The reweighting scheme corrects the bias from inferring the mass distribution alone.}
    \label{fig:validation-reweighting}
\end{figure}

\begin{figure}
    \centering
    \includegraphics[width=0.5\textwidth]{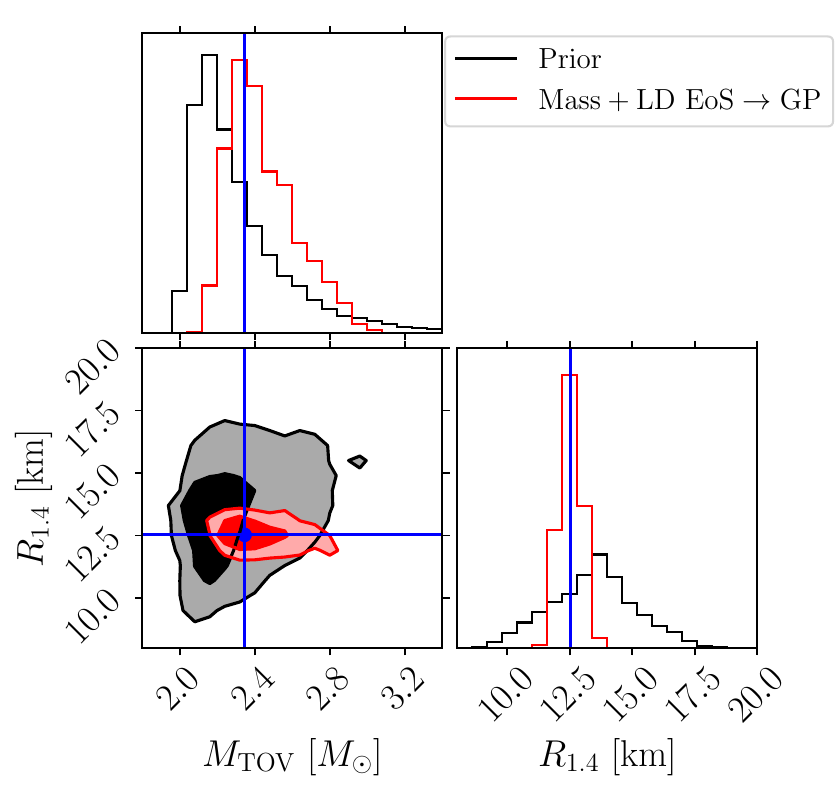}
    \caption{One- and two-dimensional posteriors for recovered EoS properties $M_{\rm TOV}$ and $R_{1.4}$ from 23 simulated BNSs.
    We plot the prior (black) and the result from reweighting to a full mass-EoS joint inference with the GP EoS model (red). The reweighting method is able to recover the true EoS (blue).}
    \label{fig:validation-reweighting-eos}
\end{figure}

We demonstrate the validity of the reweighting scheme described in Appendix~\ref{app:likelihood} with simulated GW data. 
We simulate BNS observations from a uniform mass distribution with $\alpha=0$ between $1\,M_\odot$ and $M_{\rm pop,GW} = 2.25\,M_\odot$, assigning positions and orientations isotropically, and distances according to a merger rate uniform in the frame of the source across redshifts.
Spins are distributed isotropically with uniform magnitudes up to 0.05.
Tidal deformabilities are simulated according to a pre-selected EoS from the GP prior with $M_{\rm TOV} = 2.34\, M_{\odot}$ and $R_{1.4}=12.5\,$km. 
After filtering for events that pass a detectability threshold of signal-to-noise ratio above $8$, we obtain posterior samples using \texttt{bilby}~\cite{Ashton2019}.
We then follow the procedure of Appendix~\ref{app:likelihood} to compute the joint posterior for the mass distribution and the EoS.

In Fig.~\ref{fig:validation-reweighting} we show the inferred population hyperparameters under three analyses.
The first (black) models only the mass distribution, which effectively means that the EoS model defaults to the tidal deformability prior used during sampling. 
This is selected to be uninformative to avoid restricting the posterior: flat between 0 and $1.5 \times 10^{3}$.
Since this is not in reality how the tidal deformabilities of the analyzed objects are distributed, i.e., the follow a single EoS, mass inference is slightly biased~\cite{Golomb:2021tll}.
The second analysis (green) corresponds to Eq.~\eqref{eq:proposal-marginal} that infers the mass distribution together with the lower-dimensional EoS model of Appendix~\ref{app:EoS-simple}.
The inclusion of even this simple EoS model in the inference reduces the bias compared to the true parameters.
This posterior is then used as a proposal to reweight to the final mass-EoS inference with the GP EoS model (red), which again agrees with the injected values.
Figure~\ref{fig:validation-reweighting-eos} further shows that this procedure can infer the EoS parameters.

\section{Effect of NICER observations}
\label{sec:no-nicer}

\begin{figure}
    \centering
    \includegraphics[width=.51\textwidth]{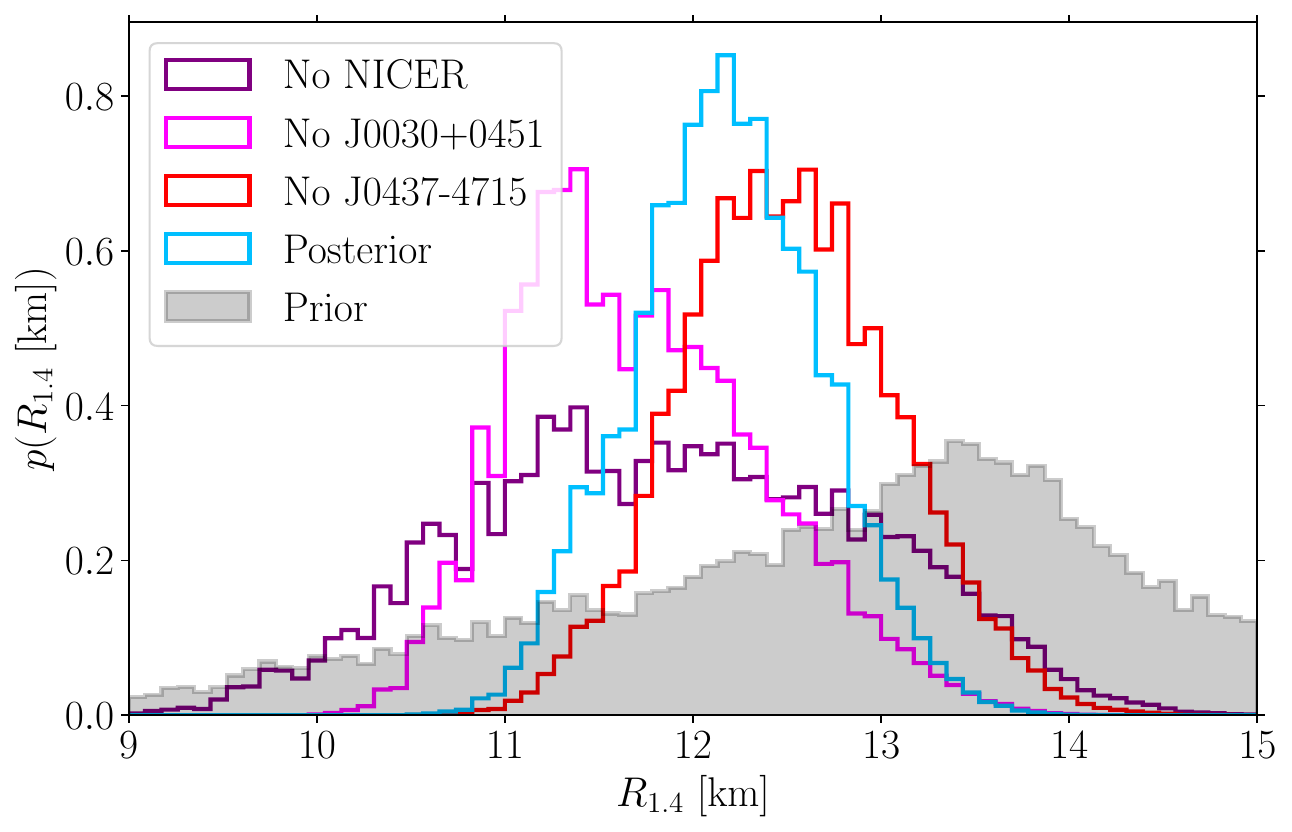}
    \caption{The effect of NICER constraints on EoS inference. We plot the prior (grey) and posterior for $R_{1.4}$, the radius of a $1.4\, M_\odot$ NS with different subsets of NICER data: all 3 pulsars (blue; main text analysis), excluding J0030+0451 (pink), excluding J0437-4715 (red), and excluding all NICER observations (purple).}
    \label{fig:R14vsnoNICER}
\end{figure}

In this Appendix we quantify the impact of NICER observations on our inference. 
Specifically, we study the impact of J0030+0451 for which there is no concurrent radio-based mass measurement and the hotspot model has a large impact on inference~\cite{Vinciguerra:2023qxq} and J0437-4715 for which only one independent analysis is available~\cite{Choudhury:2024xbk}. 
We show results for $R_{1.4}$ in Fig.~\ref{fig:R14vsnoNICER}. 
Removing any NICER pulsars leads to an increased uncertainty and a shift to lower radii (when removing J0030+0451) or larger radii (when removing J0437-4715).
However, all results are consistent with each other at the 90\% credible level.
Using no NICER data leads to $R_{1.4} = \nonicerRtyp\,$km, no J0030+0451 data to $R_{1.4} =\nojzerozeroRtyp\,$km, and no J0437-4715 data to $R_{1.4}= \nojzerofourRtyp\,$km.

\section{Uniform pulsar population}
\label{app:uniform_em_population}

\begin{figure}
    \centering
    \includegraphics[width=0.5\textwidth]{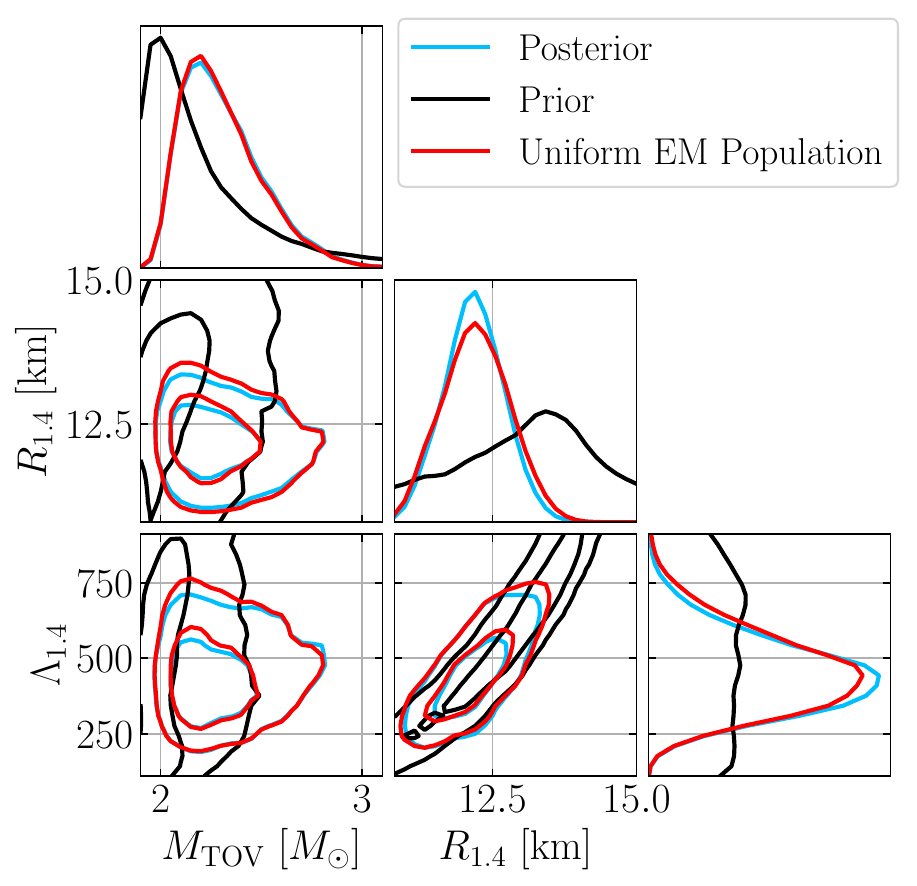}
    \caption{Impact of the EM population mass modeling on EoS inference.
    We plot the prior (black), the posterior from the full analysis (blue; same as Fig.~\ref{fig:eos-properties-corner}), and the posterior when the EM mass distribution is uniform and independnet of the EoS for J0030+0451 and J0437-4571 and uniform up to the TOV maximum mass of the EoS for J0740+6620 and J0348+0432.
    The posteriors are similar.}
    \label{fig:fixed-pop-comparison}
\end{figure}

Since selection effects for pulsar radio surveys are not well quantified, it is not clear how the observed distribution of NS masses differs from the true distribution. 
To examine the impact of the observed EM population inference, we repeat the analysis using the approach of Ref.~\cite{Legred:2021} for the EM population: it depends only on the EoS, and not on additional population hyperparameters.
The GW population is still modeled with a truncated powerlaw per Sec.~\ref{sec:gw-observations}.
We neglect all pulsars that do not contribute directly to the EoS (due to low mass) as well as spider pulsars for consistency with Ref.~\cite{Legred:2021}.
The EM data now include only J0030+0451 and J0437-4715~\cite{Miller:2019cac, Choudhury:2024xbk} with a uniform mass distribution in $[1.0{-}1.9]\,M_{\odot}$, and J0740+6620 and J0438+0432~\cite{Miller:2021qha, Antoniadis:2013pzd} with a uniform mass distribution in $[1.0{-}M_{\rm TOV}]\,M_{\odot}$, with $M_{\rm TOV}$ given by the EoS model.
This choice corresponds to a uniform distribution up to the maximum mass allowed by the EoS.
Because of this choice, EoSs that predict a larger TOV mass are penalized by an \emph{Occam} penalty for the two high-mass pulsars.   

Results are shown in Fig.~\ref{fig:fixed-pop-comparison}, where we find small changes to the inferred EoS quantities.  
In particular, $M_{\rm TOV}$ is relatively unchanged, \result{$M_{\rm TOV} = 2.27^{+0.41}_{-0.20}\,M_{\odot}$} under the fixed population, which we attribute to the cancellation of two effects.
One the one hand, the Occam penalty favors lower values of $M_{\rm TOV}$ under a fixed population.
On the other hand, under the fixed-population scheme, the mass of the heaviest pulsars is not informed by lower-mass pulsars, and therefore ends up higher, which in turn results in a higher $M_{\rm TOV}$.  
The effect of the Occam penalty and the population-informed mass estimates in practice cancel out. 
The radius and tidal deformability change somewhat more, \result{$R_{1.4} = 12.2^{+0.9}_{-1.0}\,$km}, with a ${\sim}$10\% larger uncertainty than the inferred-population case, and $\Lambda_{1.4} = \result{450^{+247}_{-175}}$ being slightly larger than the inferred-population case.  

Overall, inferring the EM mass distribution leads to marginally higher $M_{\rm TOV}$ and lower $R_{1.4}$.
Put differently, the high-density EoS is marginally stiffer and the low-density EoS is marginally softer. 
As a consequence, the maximum sound-speed is higher in order to connect the soft(er) low-density EoS to a stiff(er) high-density EoS.
This leads to increased support for violation of the conformal limit, $c_s^2 > 1/3$.  
The natural logarithm of the Bayes factor in favor of conformal violation is $\ln {\cal B}^{c_s^2>1/3}_{c_s^2 < 1/3}  = \result{ 5.85 \pm 0.30}$ for the fixed population model, and
$\ln {\cal B}^{c_s^2>1/3}_{c_s^2 < 1/3}  = \result{7.39 \pm 0.52}$ when the mass distribution of EM pulsars is inferred.

\section{Low spin assumption for GW190425}
\label{app:lowspin}

\begin{figure}
    \centering
    \includegraphics[width=0.51\textwidth]{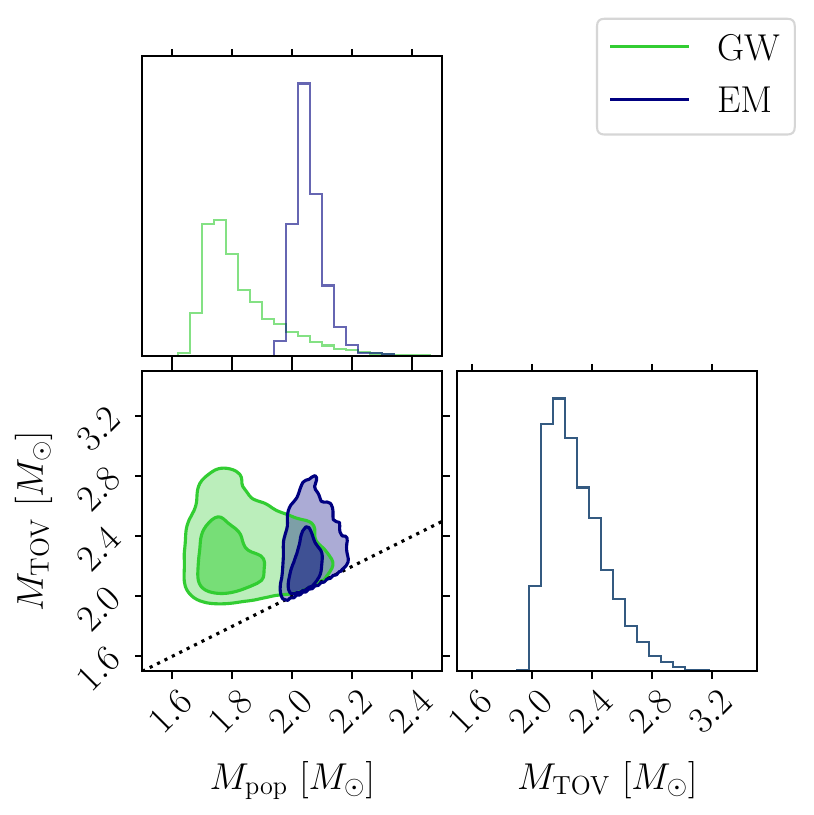}
    \caption{Similar to Fig.~\ref{fig:mpopvsmtov} but with a low-spin assumption for GW190425 of $<0.05$.}
    \label{fig:lowspin_mpopvsmtov}
\end{figure}

Assumptions about the spin of GW190425 have an effect on the inferred component masses~\cite{gw190425}. 
In the main text, we assume that the NSs in GW190425 can have dimensionless spin magnitudes up to $0.4$. 
However, other studies assume NSs have spins $0.05$, motivated by the spin distribution of pulsars in binary systems expected to merge within a Hubble time~\citep{Zhu:2017znf}. 
In Fig.~\ref{fig:lowspin_mpopvsmtov}, we present results with a low-spin assumption for GW190425, enforcing the same assumption in the sensitivity estimates as well. 
We find $M_{\rm TOV} = 2.26^{+0.39}_{-0.21}\, M_\odot$ and $M_{\rm pop,GW} = 1.79^{+0.32}_{-0.1} \,M_\odot$. As GW190425 is not the main observation informing $M_{\rm TOV}$, it values is consistent with the main analysis. 
However, as the low-spin restriction lowers the estimated masses of GW190425 due to the mass-spin correlation, we obtain a lower value for $M_{\rm pop,GW}$, though still consistent with $M_{\rm TOV}$.

\bibliography{main.bib}

\end{document}